\documentclass[a4paper,12pt]{article}
\usepackage[paper=a4paper,left=25mm,right=25mm,top=25mm,bottom=25mm]{geometry} 
\pdfoutput=1

 \usepackage{setspace}
\usepackage{authblk}

\usepackage{fullpage}
\usepackage{parskip}
\usepackage{titlesec}
\usepackage[section]{placeins}
\usepackage[dvipsnames, table]{xcolor}
\usepackage{enumitem}

\usepackage{breakcites}
\usepackage{lineno}
\usepackage{hyphenat}
\usepackage[all]{nowidow}
\usepackage{longtable}

\usepackage{centernot} 
\usepackage{lscape}

\PassOptionsToPackage{hyphens}{url}
\usepackage{hyperref} 
\hypersetup{colorlinks = true}

\hypersetup{
  linkcolor  = RedViolet,
  citecolor = Blue!90!black,
  urlcolor   = Aquamarine!80!black,
  colorlinks = true
}

\usepackage{etoolbox}
\usepackage{natbib}
\renewenvironment{abstract}
  {{\bfseries\noindent{\abstractname}\par\nobreak}\footnotesize}
  {\bigskip}
\titlespacing{\section}{0pt}{*3}{*1}
\titlespacing{\subsection}{0pt}{*2}{*0.5}
\titlespacing{\subsubsection}{0pt}{*1.5}{0pt}

\usepackage{graphicx}
\usepackage[space]{grffile}
\usepackage{latexsym}
\usepackage{textcomp}
\usepackage{tabulary}
\usepackage{booktabs,array,multirow}
\usepackage{amsfonts,amsmath,amssymb}
\usepackage{subcaption}
\usepackage{booktabs} 
\usepackage{tabularx}
\usepackage{amssymb}
\usepackage{tabularx}
\usepackage{pifont}
\usepackage{caption} 
\usepackage{graphicx}
\usepackage{hyperref}
\usepackage{comment}
\providecommand\citet{\cite}
\providecommand\citep{\cite}

\newif\iflatexml\latexmlfalse

\usepackage{comment}
\usepackage{booktabs}
\usepackage{enumitem}
\usepackage{multirow}
\usepackage{tabularx}
\usepackage{rotating}
\usepackage{wrapfig}
\usepackage{graphicx}
\usepackage{threeparttable}

\AtBeginDocument{\DeclareGraphicsExtensions{.pdf,.PDF,.eps,.EPS,.png,.PNG,.tif,.TIF,.jpg,.JPG,.jpeg,.JPEG}}

\usepackage[utf8]{inputenc}
\usepackage[english]{babel}

\usepackage{orcidlink}
\newcommand{\beginsupplement}{
        \setcounter{section}{0} 
        \renewcommand{\thesection}{S\arabic{section}}

        \setcounter{table}{0}
        \renewcommand{\thetable}{S\arabic{table}}
        \setcounter{figure}{0}
        \renewcommand{\thefigure}{S\arabic{figure}}
     }

     \newcolumntype{Y}[1]{>{\centering\arraybackslash}p{#1}}

\newcommand{\tool}[1]{\texttt{#1}}

\newcommand{\newcrossmark}{\scalebox{1}[1]{$\times$}}

\definecolor{bleudefrance}{rgb}{0.19, 0.55, 0.91}

\usepackage{array}
\newcolumntype{Y}[1]{>{\centering\arraybackslash}p{#1}}

\begin{document}
\title{To tweak or not to tweak. How exploiting flexibilities in gene set analysis leads to over-optimism.}

\def\correspondingauthor{\footnote{Corresponding author, e-mail: \href{mailto:milena.wuensch@ibe.med.uni-muenchen.de}{milena.wuensch@ibe.med.uni-muenchen.de}, Institute for Medical Information Processing, Biometry, and Epidemiology, LMU Munich, Marchioninistr. 15, D-81377, Munich, Germany.}}
\author[1,2]{Milena Wünsch \correspondingauthor{} \orcidlink{0009-0001-1982-9260}}
\author[1,2]{Christina Sauer \orcidlink{0000-0003-2425-7858}}
\author[1,2]{Moritz Herrmann \orcidlink{: 0000-0002-4893-5812}}
\author[3]{Ludwig Christian Hinske \orcidlink{0000-0001-7273-5899}}
\author[1,2]{Anne-Laure Boulesteix \orcidlink{0000-0002-2729-0947}}
\affil[1]{Institute for Medical Information Processing, Biometry, and Epidemiology, LMU Munich, Munich (Germany)}
\affil[2]{Munich Center for Machine Learning, Munich (Germany)}
\affil[3]{Institute for Digital Medicine, University Hospital of Augsburg, Augsburg (Germany)}
\vspace{-1em}
  \date{\today}
\maketitle


\begin{abstract}
Gene set analysis, a popular approach for analysing high-throughput gene expression data, aims to identify sets of genes that show enriched expression patterns between two conditions. In addition to the multitude of methods available for this task, users are typically left with many options when creating the required input and specifying the internal parameters of the chosen method. This flexibility can lead to uncertainty about the `right' choice, further reinforced by a lack of evidence-based guidance. Especially when their statistical experience is scarce, this uncertainty might entice users to produce preferable results using a `trial-and-error' approach. While it may seem unproblematic at first glance, this practice can be viewed as a form of `cherry-picking' and cause an optimistic bias, rendering the results non-replicable on independent data. After this problem has attracted a lot of attention in the context of classical hypothesis testing, we now aim to raise awareness of such over-optimism in the different and more complex context of gene set analyses. We mimic a hypothetical researcher who systematically selects the analysis variants yielding their preferred results, thereby considering three distinct goals they might pursue. Using a selection of popular gene set analysis methods, we tweak the results in this way for two frequently used benchmark gene expression data sets. Our study indicates that the potential for over-optimism is particularly high for a group of methods frequently used despite being commonly criticised. We conclude by providing practical recommendations to counter over-optimism in research findings in gene set analysis and beyond.

\end{abstract}

\sloppy

\section{Introduction}

When performing GSA, a researcher must decide on a suitable analysis strategy, including all analytical choices concerning the method, its internal parameter setting, and the preprocessing approach used to format the gene expression data as required by the selected method. The particular difficulty in this decision lies in the great multiplicity the researcher faces in all three aspects. Generally, the multiplicity of possible data analysis (and data collection) strategies is also referred to as \textit{researchers' degrees of freedom} \citep{Simmons2011}. A thorough investigation of these researchers' degrees of freedom in GSA with a focus on analysing the data is provided in our previous work \citep{wuensch2023rna}, in which we observe that there is little guidance on most of them. This leads to a considerable \textit{uncertainty} about the `right' or most suitable analysis strategy from the multitude of available options. \\
In genomics and related fields, new research findings often heavily rely on gene set analyses \citep{ballouz2017using}. Researchers might thus be tempted to exploit this uncertainty, i.e. to choose the analysis strategy that yields the `best' or most promising results after trying out several different analysis strategies concerning the choice of the GSA method, data preprocessing approach, and internal parameters. Especially researchers with little statistical experience are often unaware that such tweaking, which appears unproblematic at first view, is a form of the questionable research practice known as \textit{cherry-picking}. If they \textit{selectively report} the chosen analysis strategy and corresponding results while withholding the remaining `worse' results, the research findings are likely to be optimistically biased. As such, they may be poorly replicable based on new and independent data. \\
Note that the definition of the term `replicable' differs considerably across scientific fields and even between scientists and publications within the fields. See \citet{nosek2020replication} for a broad epistemological discussion of the concept. In the context of this study, we define replication as the attempt to recreate a previously obtained research finding by applying the same methods as in the original study on independent data. A research finding (or a whole study) is considered as `replicable' if the replication attempt is successful---where success can be defined in various ways depending on the substantive context and the considered data analysis methods. See, for example, the criteria proposed by \cite{held2022assessment} in the context of classical statistical testing.\\ 
Failure to replicate may have various reasons. The finding of the original study may be simply a type-1 error (in case a test is used) or an inflated effect in the absence of methodological flaw, or may result from flaws in design, implementation or analysis \citep{open2012open}. 
Even if flaws are not always involved in the lack of replicability of research findings, replicability is considered a core quality that all empirical research findings should fulfill. As claimed by \cite{popper2005logic}, non-replicable single findings are of `no significance to science'. \\
The statistical mechanisms behind the lack of replicability of research findings are well understood in the context of classical statistical testing. \citet{ioannidis2005most} outlines that results are often presented as `conclusive' even if they were derived from only a single (potentially flawed) study, and that flexibility in the design or the analytical mode can enable researchers to transform results from negative to positive, i.e. from non-significant to significant. It then does not come as a surprise that such results are not confirmed in later replication studies. \\
However, the lack of replicability of data analysis results has alarmingly drawn little attention beyond the context of statistical testing until very recently. A contribution to this topic is given in \cite{ullmann2023over}, who demonstrate the mechanisms of cherry-picking and its quantitative impact in terms of replicability in the specific context of unsupervised (clustering and network) analysis of microbiome data.  
In a study investigating the variability of the results generated across 13 popular GSA methods, \citet{maleki2019measuring} observe that the number of gene sets detected as differentially enriched differs by up to two orders of magnitude between the methods. This suggests that GSA may potentially be similarly subject to cherry-picking mechanisms, although in a different, perhaps less decipherable manner than classical significance testing or unsupervised analysis. 
An assessment of the impact of cherry-picking in GSA taking all types of degrees of freedom into account is still pending. The present study aims to fill this gap.
\\
More precisely, we quantitatively illustrate the questionable research practices that lead to over-optimistic (and therefore non-replicable) results in the context of GSA using real gene expression data sets. In our study, we imitate hypothetical researchers tweaking the GSA results by exploiting the inherent uncertainty about the analytical choices. We thereby proceed in a {\it stepwise} manner, reflecting the typical approach of researchers who are unaware of the impact of cherry-picking yet fundamentally well-intentioned. Using real gene expression data sets, we mimic their search for the `best' results across a wide variety of analytical choices for seven popular GSA methods, considering successively three different goals they might pursue. In particular, we investigate settings in which no gene sets are expected to be detected as differentially enriched so that the achievement of any statistically significant results through the modification of the analysis strategy can be directly interpreted as over-optimism. \\
This paper is structured as follows. We elaborate on the connection between the inherent uncertainty in the choice of the analysis strategy in GSA and over-optimism in Section~2. In Section~3, we describe the design of our study to assess the potential of GSA to generate over-optimistic and therefore non-replicable research findings, followed by the results in Section~4. Finally, we provide a discussion together with guidance to prevent over-optimism in Section~5.
\section{From uncertainty to over-optimism in gene set analysis}\label{sec2}

When selecting an appropriate analysis strategy to perform GSA, a researcher faces a noteworthy number of choices. This can lead to a considerable \textit{uncertainty} about the `right' (i.e. most suitable) choice among the corresponding options. Note that this uncertainty is to be distinguished from the uncertainty about \textit{which practical steps are generally necessary} when carrying out GSA. In our study, we assume that the user knows which steps are required to run GSA, but in each of these steps, they face a variety of options that lead to uncertainty about the right choice. \\ 
In the following, we are guided by the work of \cite{hoffmann2021multiplicity} who provide a framework of common sources of uncertainty in the general context of data analyses. We thereby focus on the four epistemic sources of uncertainty resulting from a lack of knowledge about the right strategy to \textit{analyse} the data, namely method uncertainty, model uncertainty, data preprocessing uncertainty, and parameter uncertainty. We translate this general framework to the context of GSA to outline the choices a researcher is confronted with. See Table~\ref{overview_uncertainties} for an overview. Thereby, we assume that the data generation has been completed and the (raw) gene expression data set is available. Note that there are additional sources of uncertainty anchored in the \textit{generation} of the gene expression data set which can also lead to variability in the results even if the analysis strategy is fixed. While we do not include these sources of uncertainty in our analysis, we address them briefly in Section~1 in the supplement. \\
In the context of GSA, \textit{method uncertainty} refers to the uncertainty about the choice of a method to investigate differential enrichment of the gene sets between the conditions. The wide variety of available methods from which a researcher has to choose becomes clear when inspecting the comprehensive reference database on GSA methods by \citet{xie2021popularity}. This database contains around 150 GSA methods assigned to Over-Representation Analysis (ORA) and Functional Class Scoring (FCS) alone. However, there is little guidance on how to make a suitable choice. \citet{ballouz2017using} claim that there is even no general consensus on how to benchmark the available methods to derive such guidance. Furthermore, \citet{xie2021popularity} make a three-fold observation. Firstly, each benchmark study typically compares only a small subset of all available methods, meaning that the performance of the majority of methods has not even been investigated beyond the original papers that introduced them. Secondly, the benchmark studies often contradict each other in their results regarding the best and poorest performers, resulting in some methods simultaneously occupying the top and bottom positions in different performance rankings. Lastly, there appears to be a discrepancy between the performance of these methods and their popularity among the users. These observations underline that the right choice of a GSA method is far from clear in practice. \\
Of important note, we use the general term \textit{method} to refer to both theoretical and computational methods. By \textit{theoretical} method we mean the method's general concept and features as typically described textually in an original scientific article. In contrast, we refer to their practical implementations in the form of web-based applications or software packages as \textit{computational} methods (and use typewriter font for their corresponding name, i.e. `\tool{method}'). For instance, the \textit{theoretical} method `Gene Set Enrichment Analysis' from the publication of \citet{subramanian2005gene} is implemented in several \textit{computational} methods.  While the computational method \tool{GSEA} \citep{subramanian2005gene, mootha2003pgc}, which is a web-based application, exactly implements Gene Set Enrichment Analysis as introduced in the original paper, the user can also choose from computational methods that implement variations of it, such as \tool{GSEAPreranked} and GSEA provided by the \texttt{R} package \texttt{clusterProfiler} \citep{wu2021clusterprofiler}. In the remainder of this paper, we will use the generic terms `method' and `method uncertainty' and not further address the distinction between theoretical and computational methods. \\
In the context of GSA, \textit{model uncertainty}, which arises from the uncertainty about how to adequately model the underlying system, is implicitly included in method uncertainty. Imagine, for instance, that a researcher chooses between the two popular methods DAVID and Gene set Enrichment Analysis. This implies the choice between the general approaches ORA and FCS and their assumptions on the underlying biological system on which the corresponding methods are based. For instance, while ORA methods typically assume a hypergeometric distribution as the underlying null distribution, FCS methods assess differential enrichment non-parametrically. \\
No less pronounced than method uncertainty (and the implied model uncertainty) is the uncertainty about how to process the gene expression data into the format required by the chosen method (\textit{data preprocessing uncertainty}). In earlier work, we have observed that this aspect is often neglected in practical applications of GSA as well as user manuals provided alongside the corresponding methods \citep{wuensch2023rna}. This, again, results in little guidance for researchers. One of many examples is the choice of a method for differential expression analysis to generate the required input for ORA methods, for which the popular methods limma \citep{law2014voom}, DESeq2 \citep{love2014moderated}, and edgeR \citep{robinson2010edger} are only a small selection of all available options.\\
Finally, the researcher is confronted with uncertainty about the choice of parameter values within the chosen GSA method (\textit{parameter uncertainty}), arising from existing flexibility in the parameters to adapt the analysis strategy to the given research question. An example is the parameter `gene set database', for which there are a variety of options that differ in structure and additional aspects related to the modelling of the underlying biological system. Note that a loose form of guidance is available for some parameters in the form of default values, while for others, such as the gene set database, the user has to make the decision autonomously.

\begin{table}[ht]
\small
\centering
    \caption{Overview of the sources of uncertainty arising in the analysis of the gene expression data using gene set analysis.}
    \begin{tabularx}{0.96\textwidth}{p{0.17\textwidth}p{0.35\textwidth}p{0.35\textwidth}}
    \toprule
      Uncertainty source  & Description & Example\\
         \midrule
         \rowcolor{gray!25}
       Model uncertainty  & Which model describes the underlying system best? & Should I choose an ORA or an FCS method? \\
       \rowcolor{gray!7}
       Method uncertainty & Which method should I choose? & Should I choose \tool{GSEA} or \tool{DAVID}?  \\ 
       \rowcolor{gray!25}
       Data preprocessing uncertainty & Which approach should I choose to generate the input object required by the GSA method? & Which approach to pre-filtering should I choose?  \\ 
       \rowcolor{gray!7}
       Parameter uncertainty & Which values of the input parameter for the GSA method should I choose? & Should I choose Gene Ontology or KEGG as the gene set database?\\ 
        \bottomrule   
    \end{tabularx} \\ 

    \label{overview_uncertainties}
\end{table}

Combined with a lack of clear practical guidance, these uncertainties might impel users to select the GSA method, its underlying parameters (including parameters such as the gene set database), and data preprocessing approach based on which choice(s) yield(s) preferable results to their research question. Such practice may seem natural at first glance. After all, pitfalls of analysis strategies often only come to light when the analyses are run, and it is then acceptable to modify the original planned strategy. Some researchers may not realise that choosing the method that yields preferable results is more than just a reaction to unforeseen problems of the planned analysis strategy. It indeed amounts to the \textit{questionable research practice} termed \textit{cherry-picking}.  \\
\citet{hoffmann2021multiplicity} accentuate that a selective reporting of research findings generated using this `cherry-picking approach' often results in presenting over-optimistic and thus non-replicable findings including false positive test results and inflated effect sizes, as outlined in the introduction. Assuming the well-intentioned nature of researchers engaging in cherry-picking (as opposed to someone who maliciously intends to manipulate the results), we want to realistically assess the extent to which their tweaking of the GSA results in the above-described manner leads to over-optimism.

\section{Design of the study}

\begin{sidewaysfigure}
    \centering
    \includegraphics[scale =0.8]{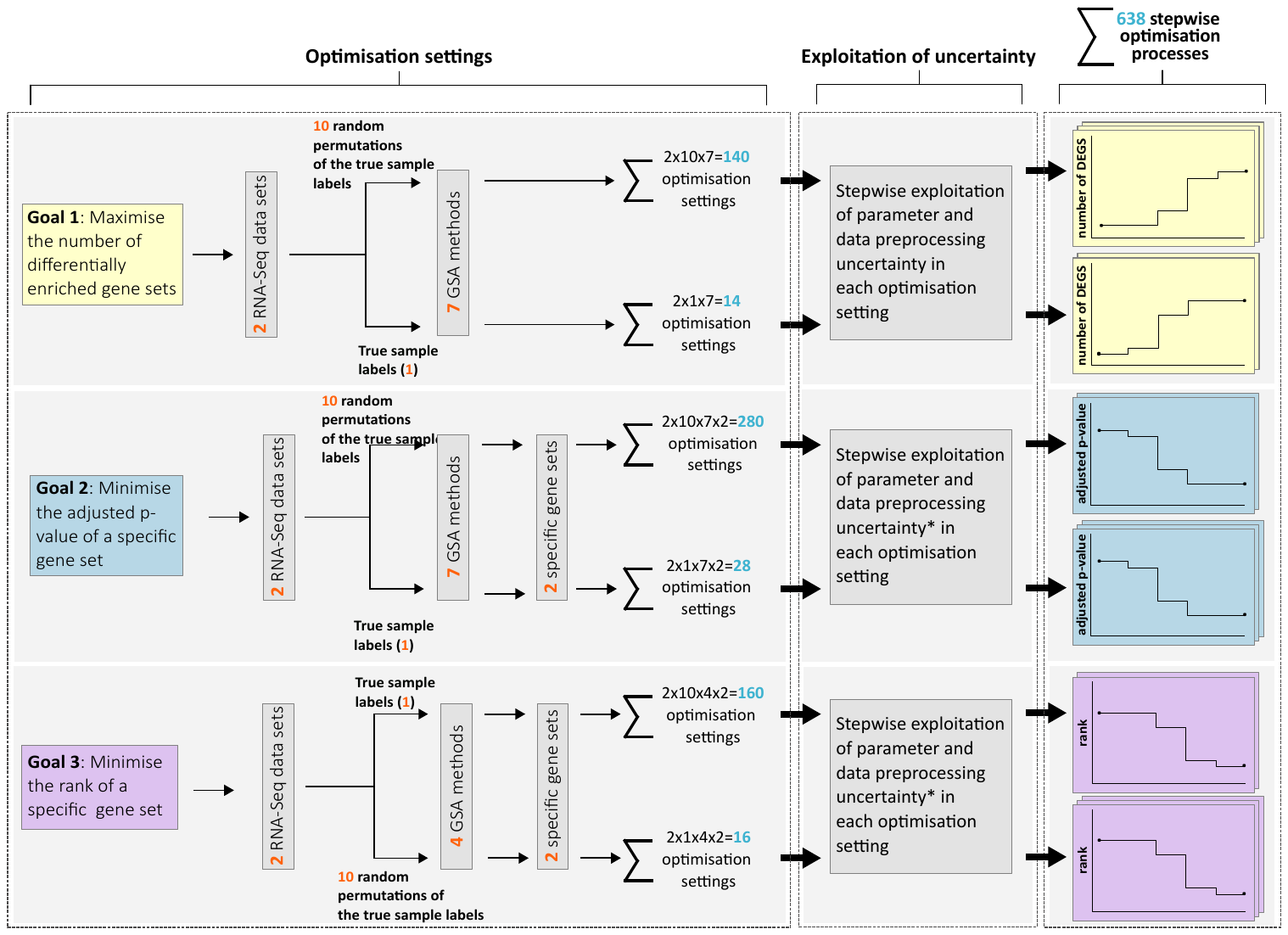}
    \caption{Overview of the study design to investigate the potential for over-optimistic results in a total of 
    $638$ optimisation settings, resulting in $638$ separate optimisation processes. The asterisk `$\ast$' refers to the fact that for goals 2 and 3, the choice gene set database cannot be exploited in the corresponding optimisation processes.}
    \label{fig_studydesign}
\end{sidewaysfigure}

The study aims to systematically assess the potential of GSA for the generation of over-optimistic and thus non-replicable results as a consequence of the exploitation of the inherent uncertainties combined with selective reporting. We expect the potential of over-optimism to vary depending on the considered GSA method, gene expression data set, and goal of the analysis. Therefore, we imitate the behaviour of a hypothetical researcher in their attempt to tweak (i.e. to \textit{optimise}) the GSA results in a variety of settings described in Section~\ref{sec_design_settings}. The exploitation of uncertainty in each fictive setting leads to a separate \textit{optimisation process}. While the details of the uncertainty exploited are presented in Section~\ref{sec_design_uncertainties}, the structure underlying each optimisation process with its different steps is described in Section \ref{sec_stepwise_opt}. For a graphical illustration of the overall study design, see Figure~\ref{fig_studydesign}.

\subsection{Settings} 
\label{sec_design_settings}
Each setting (also referred to as \textit{optimisation setting}) is defined through a unique combination of 
\begin{itemize}[noitemsep]
     \item[(i)] one (of three) optimisation goal(s) that drive(s) the optimisation, 
    \item[(ii)] one (of two) gene expression data set(s),
      \item[(iii)] one specific assignment of the conditions to the samples (either the true sample labels or one of the ten random permutations), 
    \item[(iv)] one (of seven) GSA method(s),
    \item[(v)] for two of the goals mentioned in (i): one (of two) gene set(s).
\end{itemize}
This results in a total of $638$ optimisation settings and correspondingly $638$ optimisation processes. For each optimisation process, over-optimism in the tweaked results is assessed relative to the `default' results, i.e. the results arising from the default choice in all analysis steps of GSA in which the hypothetical researcher faces uncertainty. For a description of the specification of the default choices, see Section ~\ref{sec_stepwise_opt}. 
In the remainder of this section, we elaborate on aspects (i) to (v) that define the individual optimisation settings.\\
\textit{\textbf{(i) Optimisation goals}}\\
When exploiting the uncertainty inherent to GSA in an attempt to tweak the results, a researcher typically has a specific criterion (i.e. `goal') in mind. In the following, we define three (distinct) possible goals, resulting in three different ways the hypothetical researcher tries to induce results that they consider satisfactory. \\
First, it can be assumed that a researcher would not undergo the complex and laborious procedure of generating a gene expression data set and analysing it without hoping for significant results in the first place. Furthermore, a large number of significantly enriched gene sets provides flexibility when reporting a study. One may focus on those significantly enriched gene sets that better fit the `storyline' of the paper. In the framework of our study, we translate this preference for a large number of significantly enriched gene sets by formulating optimisation goal~1 as `maximising the number of differentially enriched gene sets'. \\
The two remaining goals follow an alternative intuition. Researchers often have an explicit expectation as to which gene set constitutes an interesting research finding and therefore try to maximise its relevance in the GSA results. Relevance of a specific gene set may be defined in different ways, leading us to consider the optimisation goals 2 and 3: `minimising its adjusted $p$-value' and `minimising its rank among the remaining gene sets', respectively. We thereby assess the extent to which a user can influence the ranking of the gene sets in the GSA results to induce a significant association between the condition of interest and a particular gene set. In the context of our study mimicking a hypothetical researcher's approach, we have to choose the gene sets to be involved in goals 2 and 3, see below (v).\\
\textit{\textbf{(ii) Gene expression data sets}}\\
The three goals are separately considered for two RNA-Seq gene expression data sets. The selection criteria for the data sets, leading to two real data sets frequently used for benchmarking, can be found in Section Section~3.1 in the supplement. The first data set, in the following referred to as `Pickrell data set', contains gene expression measurements of $52580$ genes which were extracted from the lymphoblastoid cell lines of $69$ independent Nigerian individuals \citep{pickrell2010understanding}. The samples are labelled according to the sex of the individuals ($n=29$ males, \ $n=40$ females). We obtained the data set from version 1.34.0 of the \texttt{R} package \texttt{TweeDEseqCountData}  \citep{tweeDEseqCountData_package}. \\
The second data set, referred to as `Bottomly data set', was used to detect genes that are differentially expressed between the two inbred mouse strains `C57BL/6J' ($n=10$) and `DBA/2J' ($n=11$) from a total of $36536$ genes \citep{bottomly2011evaluating}. We obtained this data set from the ReCount project \citep{frazee2011recount}. \\
\textit{\textbf{(iii) Sample labels}}\\
In our study, we focus on scenarios where the ground truth is that no gene sets are differentially enriched between the conditions of interest. Any improvement of the GSA results through the exploitation of uncertainty (in the respective contexts of goals 1 to 3) can thus be interpreted as over-optimism. Given the above-described gene expression data sets, we obtain such scenarios by permuting the true sample labels randomly across the samples, thereby removing the biological meaningfulness from the data. We repeat the permutation procedure ten times, resulting in ten random permutations of the true sample labels. \\
However, in reality, over-optimism might also occur when the ground truth is truly unknown. Researchers might still aim to improve the results, for instance, to obtain a better storyline of the gene sets to report as findings. We therefore repeat our study on the true (i.e. non-permuted) sample labels. Note, however, that we cannot interpret the corresponding improvement as over-optimism exclusively. It may also be possible that the default analytical choice(s) did not model the underlying biological system adequately---and that the `optimised' choice of the analysis strategy (parameter setting, data preprocessing approach) happens to better do so. When considering data sets with the true sample labels, our focus is therefore on the quantification of the \textit{variability} in the GSA results as the consequence of parameter and data preprocessing uncertainty, rather than on over-optimism.\\
\textit{\textbf{(iv) GSA methods}}\\
We consider a selection of seven GSA methods from the reference database provided by \citet{xie2021popularity}. Six of these seven methods are chosen for their popularity, whereas the seventh is selected for its good overall performance. Note that this selection procedure results in a restriction to methods categorised as ORA or FCS. For further details on the selection process and short descriptions of the resulting GSA methods, inspect our earlier work \citep{wuensch2023rna}. An overview of the methods included in our study can be found in Table \ref{overview_methods}. \\
It is important to note that, while the choice of a GSA method (and the underlying model) is a source of uncertainty in practice, we do not exploit method uncertainty in the individual optimisation processes in our study. Instead, the optimisation for a given optimisation goal, gene expression data set, and assignment of the sample labels is performed for each GSA method separately: in each optimisation process, the GSA method is considered to be fixed. This has the advantage that we can better investigate and compare the behaviours of the methods. Method uncertainty is, however, implicitly considered when comparing the results of the optimisation processes across the GSA methods. 
Of additional note is that three methods from our selection, namely \tool{DAVID}, \tool{GSEA}, and \tool{GSEAPreranked}, are web-based applications for which all optimisation processes have to be performed manually (i.e. by hand). Time constraints lead us to limit our study for these three methods to the optimisation goals 1 and 2, whereas goal~3 is omitted.
\begin{table*}[ht]
\small
\centering
    \caption{Overview of the GSA methods included in our study.}
    \begin{tabularx}{0.85\textwidth} {p{0.2\textwidth}p{0.15\textwidth} p{0.15\textwidth}p{0.23\textwidth}}
    \toprule
      GSA method & Implemented \newline in & Selection \newline criterion  & Introduced by  \\
         \midrule
         \rowcolor{gray!25}
         \tool{GOSeq}& \texttt{R} & Popularity & \citet{young2010gene} \\
        \rowcolor{gray!7}
       \tool{DAVID} & Web &Popularity & \citet{huang2009bioinformatics,huang2009systematic} \\
       \rowcolor{gray!25}
       ORA by \newline \tool{clusterProfiler} & \texttt{R}& Popularity & \cite{wu2021clusterprofiler}\\
       \rowcolor{gray!7}
        \tool{PADOG} & \texttt{R}& Performance & \citet{tarca2013comparison}\\
        \rowcolor{gray!25}
       GSEA by \newline \tool{clusterProfiler} & \texttt{R}& Popularity & \citet{wu2021clusterprofiler}\\    
        \rowcolor{gray!7}
       \tool{GSEA} & Web & Popularity & \citet{subramanian2005gene},  \newline\citet{mootha2003pgc}\\
       \rowcolor{gray!25}
       \tool{GSEAPreranked} & Web& Popularity & \citet{subramanian2005gene},  \newline \citet{mootha2003pgc}\\   
        \bottomrule 
    \end{tabularx} \\ 

    \label{overview_methods}
\end{table*} \\
\textit{\textbf{(v) Gene sets (for goals 2 and 3)}}\\
Goals 2 and 3 refer to a specific gene set whose adjusted $p$-value or rank, respectively, is to be minimised within the optimisation process. In practice, the preferences of researchers for a specific gene set arise from previous experiences, literature, or the hypotheses they want to investigate. In the context of our study, we have to define the preferences of our hypothetical researcher in an arbitrary but plausible way. In our initial attempt to select two common gene sets for all GSA methods, we encountered obstacles that led us to consider different gene sets for the different methods. A description of the selection process alongside an overview of the selected gene sets for all methods and both gene expression data sets is provided in Section~3.1 in the supplement. For simplicity, we only refer to `gene set 1' and `gene set 2' in the rest of this paper.

\subsection{Exploited uncertainties}
\label{sec_design_uncertainties}
\begin{figure}
    \centering    \includegraphics[scale = 0.65]{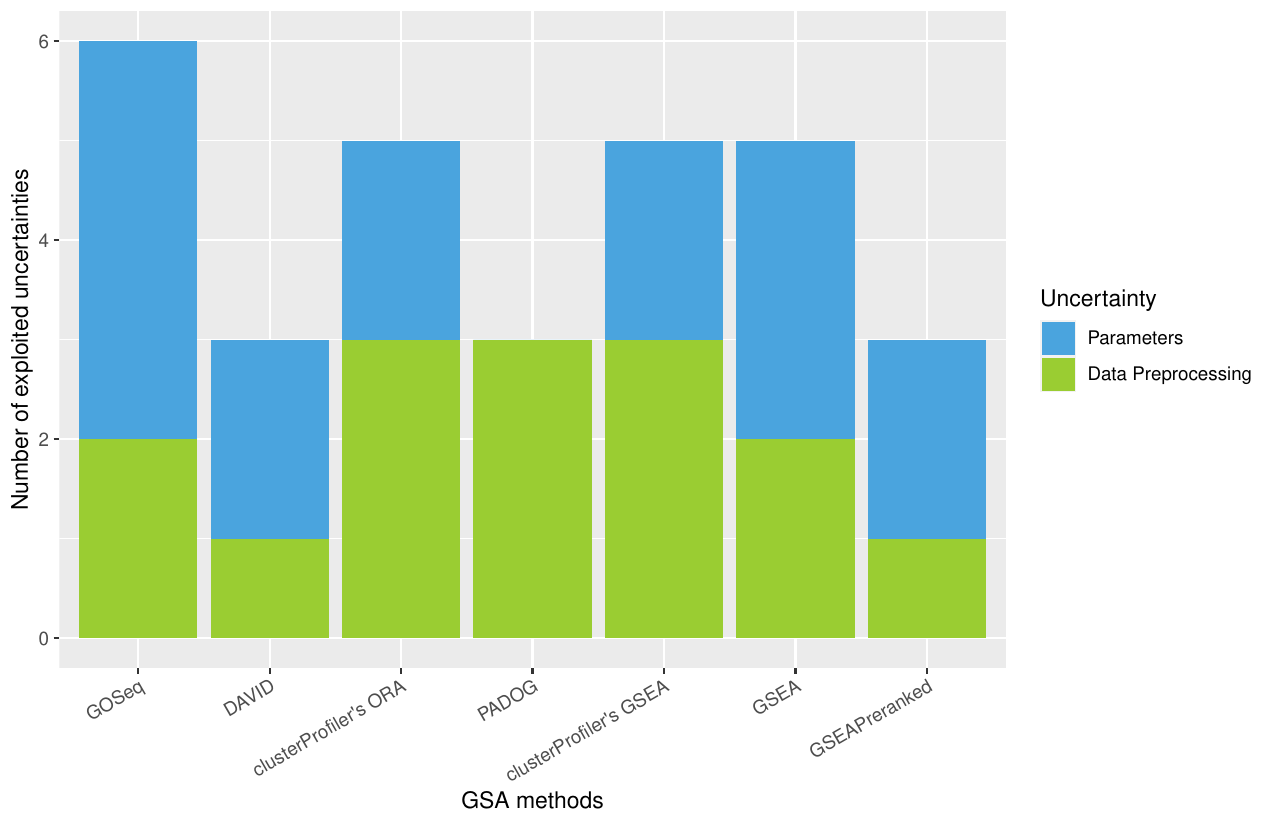}
    \caption{Overview of the number of choices affected by uncertainty that are exploited in our study for each method under investigation. The numbers are additionally split based on the type of uncertainty the corresponding choice is associated with.}
    \label{fig_flexiblechoices}
\end{figure}

An overview of the numbers of exploited data preprocessing and parameter uncertainties for each of the seven investigated GSA methods can be found in Figure \ref{fig_flexiblechoices}. This figure illustrates that the ratio between data preprocessing and parameter uncertainties can differ notably between the methods. For instance, \tool{PADOG} does not offer any flexibility in terms of the parameter setting, while for \tool{GOSeq}, four of the six exploited uncertainties arise from parameter uncertainty. Our underlying assumption that researchers are \textit{well-intentioned} implies that we deliberately renounce to exploit some uncertainties when we consider that this would amount to a willful manipulation of the results. This concerns adaptions in the GSA workflow that might be inappropriate in the given statistical or biological context or those explicitly discouraged by the author of the respective method. Furthermore, we only consider those adaptions that can be carried out without excessive effort. \\

\textit{\textbf{Data preprocessing uncertainty}} \\
We exploit the uncertainty about the approach to pre-filtering of lowly expressed genes, the removal of duplicated gene IDs as a result of gene ID conversion (while we do not exploit the choice of an approach to gene ID conversion itself), the method for differential expression analysis (`DE method') for ORA and some of the FCS methods, and the method for transformation (which typically includes normalisation) for the remaining FCS methods. Note that not all of these steps apply to each of the GSA methods from our selection since the methods often differ in the required input object. Furthermore, the options available for one step might depend on the choice made in a previous step. For instance, the approach to pre-filtering typically differs between different DE methods. For a more detailed description of the exploited uncertainties in the data preprocessing steps, including a description of the respective options, see Section~2.1 in the supplement. \\
\textit{\textbf{Parameter uncertainty}} \\
In the process of optimising the results, we exploit uncertainty about the choice of the gene set database, the so-called `universe' (for ORA), the method for the calculation of the $p$-value, as well as the gene-level statistic and the weight for FCS. Note that not all uncertainties apply to all GSA methods. For a more detailed overview, refer to Section~2.2 in the supplement. Note that we exploit the choice of the gene set database only for goal~1, i.e. when maximising the number of differentially enriched gene sets. Indeed, the specific gene sets considered for goals 2 and 3, stemming from a certain gene set database, typically do not exist in the same form in another gene set database. The reason for this is that the gene set databases can differ greatly in their structure and the gene sets they contain. 
Furthermore, for goal~1, we restrict ourselves to the gene set databases offered alongside the individual GSA methods and do not work with a customised gene set database that can be uploaded to the method.

\subsection{Stepwise optimisation process} \label{sec_stepwise_opt}

In this section, we focus on the structure underlying all optimisation processes of the GSA results. Real researchers, particularly if well-intentioned, are unlikely to try out all combinations of analytical choices affected by uncertainty. Instead, they are more likely to tweak the results in a stepwise manner---consciously or subconsciously. We also adopt a stepwise approach in our optimisation processes. This means that we exploit the uncertainties in a specific order such that the optimal choice of a specific step (in the context of the corresponding optimisation goal) is based on the optimal choices from the previous step(s). Optimisation is thus not performed globally for all uncertainties simultaneously, because we consider such an approach unrealistic in practice.

More precisely, we specify the order of the choices a priori and (roughly) in alignment with the natural order of the corresponding steps required to conduct GSA. For instance, uncertainties arising from data preprocessing uncertainty are exploited before addressing those emerging from parameter uncertainty. The intuition behind this order is that a user of GSA must perform data preprocessing before running the actual GSA method (and specifying the corresponding parameters). However, there are instances where we have to make exceptions to this order, namely when it leads to practical difficulties in the optimisation processes. For example, for ORA methods, the choice of the DE method is optimised before pre-filtering (while the natural order is reversed). The reason for this is the previously mentioned fact that different DE methods propose different pre-filtering approaches. \\
For each uncertain choice in the GSA workflow, we set a default option a priori. For those steps where a common default option exists (which is often the case for parameters), we set the default choice for our study accordingly. For those steps for which no default exists (such as the gene set database for many methods), we set the default choice for our analysis arbitrarily. Furthermore, we specify an ordering of the alternative options, which will be used as a criterion in case several alternative options yield exactly the same improvement; see below. 

Having specified the set of uncertain choices (i.e. steps), their order, the default, and valid alternative options including their order, the optimisation process proceeds as follows. 
The optimisation process starts with all choices in their default configuration (step~1). The value to be optimised (i.e. the number of differentially enriched gene sets for goal~1, the adjusted $p$-value, or rank of the considered gene set for goals 2 and 3, respectively) obtained in this default configuration is reported as the starting point (\textit{default results}). Then, in step~2, the first uncertain choice is exploited such that the default results are compared to all results stemming from the alternative options. An alternative option is then adopted as the \textit{optimal option} for this step if it leads to an improvement of the results, namely an increased number of differentially enriched gene sets (for goal~1), a decreased adjusted $p$-value of the considered gene set (for goal~2), or a decreased rank of the considered gene set among the remaining ones (for goal~3). If several alternatives lead to an improvement of the GSA results, the alternative that leads to the greatest improvement is selected. If the improvements are equally strong, the alternative option is selected according to the preliminary fixed ordering. 
In contrast, the default option for this step is retained if none of the alternative options leads to an improvement. The GSA results arising from the optimal option are then denoted as the \textit{current optimal results}. \\
In the third step, the procedure just described is repeated for the second uncertain choice. Thereby, the current optimal results from step~2 serve as the default results in this step. This procedure is carried out in the same manner for each uncertain step and in the order established previously. That way, the optimal choice in each step is based on the optimal choices and corresponding optimal results from all previous steps.\\
By examining the results optimised in this stepwise manner and comparing them to the respective default results, it is possible to assess the variability in the results generated by exploiting uncertainty. Most importantly, for the permuted sample labels, it is possible to assess the level of over-optimism induced by this optimisation process. 
\section{Results}

The presentation of the results of our study is structured according to the three optimisation goals. For each goal, the results are further split according to the assignment of the sample labels (ten permutations versus true sample labels). For each method considered for the respective goal, we then contrast the `tweaked' GSA results from the associated optimisation processes to the corresponding `default' results. This allows for a comparison between the GSA methods regarding the potential for over-optimism. \\
Since the results are generally similar for the Pickrell and the Bottomly data set, we focus on the results for the former data set, while referring to Section~4.2 the supplement for the results of the Bottomly data set. To gain a better understanding of the stepwise structure of the optimization processes using a concrete example, see Section~4.1 in the supplement. \\
For simplified readability, we abbreviate the term `sample label permutation' with `permutation' in the following.

\subsection{Results for goal~1: maximise number of differentially enriched gene sets (DEGS)}

\begin{figure}
    \centering    \includegraphics[scale = 0.57]{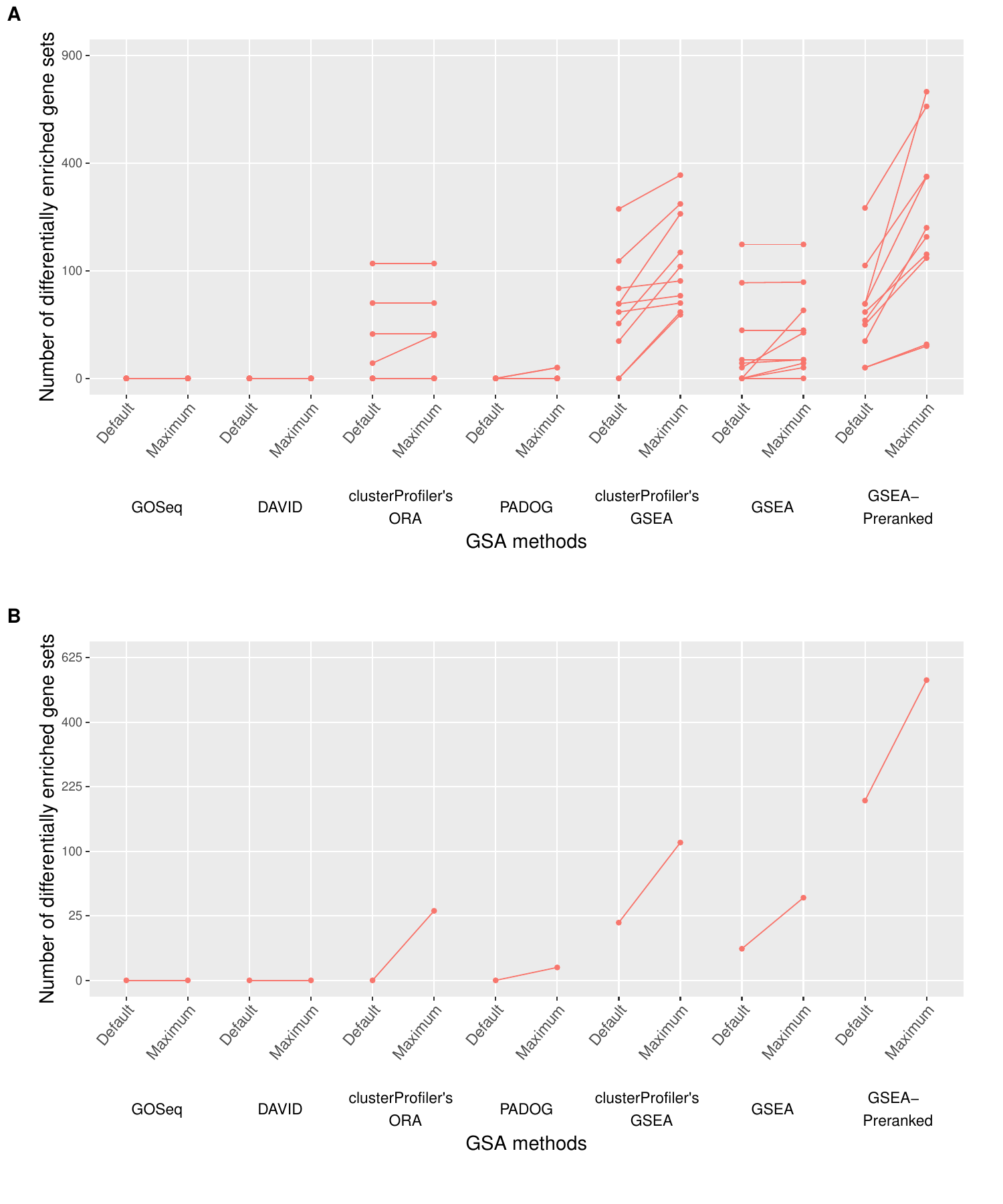}
    \caption{\small Goal~1: The optimised numbers of differentially enriched gene sets in the Pickrell data set (`Maximum'), obtained through the exploitation of uncertainty, are compared to the corresponding numbers resulting from the default analytical choices (`Default'). For each optimisation process, the associated optimised and the default number are connected through a line. (A) presents the results for the ten random permutations and (B) for the true sample labels. On the $x$-axis, the individual methods investigated in the context of goal~1 are displayed. Special attention must be paid to the transformed scale of the $y$-axis. This transformation enables visibility of small increases in the number of DEGS, while particularly large increases appear smaller than they actually are.}
    \label{fig_results_n_DEGS}
\end{figure}

For a graphical illustration of the results for goal~1, see Figure~\ref{fig_results_n_DEGS}.

\textit{\textbf{Random sample label permutations:}} 
Over-optimism concerning goal~1 mainly affects the GSEA-based methods (\tool{GSEA}, \tool{GSEAPreranked}, and \tool{clusterProfiler}'s GSEA). In particular, the number of DEGS cannot be increased for any of the permutations for \tool{GOSeq} and \tool{DAVID} and only in a small minority of permutations for \tool{clusterProfiler}'s ORA and \tool{PADOG}. For \tool{PADOG} especially, the number of DEGS does not exceed $1$ DEGS after the exploitation of uncertainty in any of the permutations. \\ 
For the web-based application \tool{GSEA}, an increase in the number of DEGS is obtained for just over half of the permutations, in three of which the initial number of $0$ can be tweaked to a non-zero one. For instance, we observe an increase from $0$ to $40$ DEGS in one permutation solely through the specification of an alternative weighting pattern of the genes in the computation of the enrichment score. Note that for the web-based method \tool{GSEA}, the set of analytical choices leading to a tweak in the corresponding results differs greatly between the permutations.  \\ 
The observations for the remaining (GSEA-based) methods \tool{GSEAPreranked} and \tool{clusterProfiler}'s GSEA are particularly striking. Even before exploiting any uncertainty, the initial numbers of DEGS considerably exceed $0$ in the vast majority of permutations. Note that this observation cannot be viewed as over-optimism in the sense considered in this paper. However, it provides information about the general reliability of these GSA methods. In particular, it coincides with the observation that FCS methods that require as input an already ranked list of the genes (as opposed to generating the ranking internally) have inflated false discovery rates \citep{maleki2020gene, wu2012camera}.  \\
Furthermore, an increase in the number of DEGS can be observed with these two methods for all permutations, amounting to several magnitudes in the great majority of cases. Thereby, a striking pattern as to which analytical choices in the data preprocessing and parameters trigger an increased number of DEGS is particularly visible for \tool{GSEAPreranked}. Consisting of the choice of the DE method and the assignment of equal weight to all genes in the computation of the enrichment score, the increase from $12$ to $196$ DEGS in one permutation is only one of many examples where it leads to over-optimistic results. \\
According to the user manual provided alongside \tool{GSEAPreranked}, the option of assigning equal weight to each gene in the computation of the enrichment score can be viewed as a `conservative scoring approach'. It is recommended over the default of weighting each gene by its absolute value of the gene-level statistic when unsure about the biological meaningfulness of the magnitude of the ranking metric for the user's research question \citep{subramanian2005gene, mootha2003pgc}. A user could thus easily justify the exploitation of this uncertainty with the recommendations from the user manual. \\
For \tool{clusterProfiler}'s GSEA, the set of uncertain choices leading to the highest increase in the number of DEGS varies more strongly between the permutations. Nevertheless, we also observe notable increases for this method, such as from $0$ to $38$ DEGS through the choice of the DE method and the pre-filtering approach. \\
\textit{\textbf{True sample labels:}} 
For the true sample labels, an increase in the number of DEGS by exploiting data preprocessing and parameter uncertainty is achieved for all GSA methods apart from \tool{GOSeq} and \tool{DAVID}. For the latter two, the number of DEGS amounts to $0$ before and after the exploitation of uncertainty. For \tool{PADOG}, a moderate increase from $0$ to $1$ DEGS is achieved through the choice of pre-filtering. In contrast, for \tool{GSEAPreranked}, the initial number of DEGS, amounting to almost $200$, is already exceptionally high and can even be further doubled through the choice of the DE method as part of data preprocessing. This choice proves to be a trigger for increase not only for \tool{GSEAPreranked}. While for \tool{clusterProfiler}'s ORA, the number of DEGS is thus increased from $0$ to $26$, the initial number of $20$ DEGS is further raised by factor six through the modification of the DE method for \tool{clusterProfiler}'s GSEA.

\subsection{Results for goal~2: minimise adjusted \texorpdfstring{$p$} \ -value of a specific gene set}
\begin{figure}
    \centering    \includegraphics[scale = 0.56]{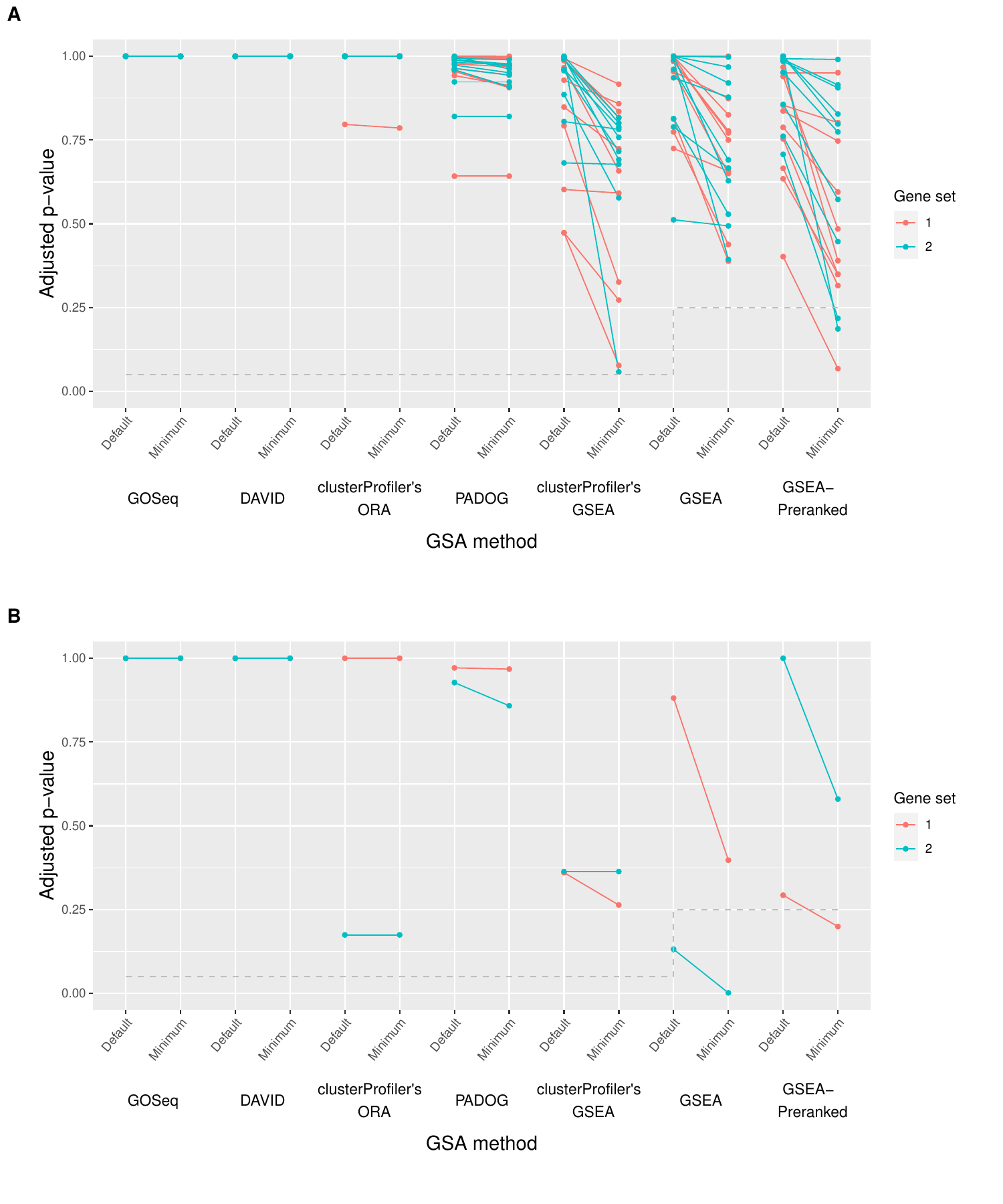}
    \caption{\small Goal~2: The optimised adjusted $p$-values (/$q$-values) in the Pickrell data set (`Minimum'), obtained through the exploitation of uncertainty, are compared to the corresponding values resulting from the default analytical choices (`Default'). Note that for the web-based applications \tool{GSEA} and \tool{GSEAPreranked}, the $q$-value is used to assess differential enrichment instead of the adjusted $p$-value. For each optimisation process, the associated optimised and the default adjusted $p$-value (/$q$-value) are connected through a line. (A) presents the results for the ten random permutations and (B) for the true sample labels. On the $x$-axis, the individual methods investigated in the context of goal~2 are displayed. The results for gene set 1 are shown in red and those for gene set 2 in blue. The dashed grey line indicates the significance threshold for each method below which a gene set is considered differentially enriched.}
    \label{fig_results_padj}
\end{figure}

Note that the significance threshold considered by the web-based applications \tool{GSEA} and \tool{GSEAPreranked} is a $q$-value of $<0.25$; see \citet{storey2002direct} for a definition of the $q$-value. The remaining methods detect a gene set as differentially enriched if its $p$-value adjusted using the Benjamini-Hochberg procedure is lower than $0.05$. We stick to these default settings in our interpretation of the analyses of goal~2. See Figure~\ref{fig_results_padj} for a graphical illustration of the results for goal~2. \\
\textit{\textbf{Random sample label permutations:}} 
Analogous to goal~1, the degree of over-optimism regarding goal~2 is the highest for the GSEA-based methods. Before and especially after exploiting uncertainty, these three methods generally indicate lower adjusted $p$-values (/$q$-values), respectively, compared to the remaining methods. \\
Similar to goal~1, the adjusted $p$-values of the respective gene sets cannot be tweaked for \tool{GOSeq} and \tool{DAVID} (all adjusted $p$-values are equal to 1 even after optimisation). Similar observations are made for \tool{clusterProfiler}'s ORA and \tool{PADOG}, where decreases in the adjusted $p$-value through the exploitation of data preprocessing and parameter uncertainty are, if at all existent, negligible (i.e. less than 0.02 in the vast majority of permutations). \\
 For the web-based application \tool{GSEA}, we observe moderate to notable decreases in the $q$-value in the vast majority of permutations. However, none of the optimisations applied in the analysis of the permuted data turns an initially non-significant $q$-value into a significant one.  \\
In contrast, there are three permutations for \tool{GSEAPreranked} in which the tweaking of the GSA results leads to the detection of differential enrichment of the respective gene set that was initially not found to be enriched. In particular, we observe a $q$-value decrease from $1$ to $0.19$ in one permutation. As can also be observed in the majority of the remaining permutations, this strong decrease is triggered by the choice of the DE method and the modification of the weighting pattern of the genes in the computation of the enrichment score. \\
For \tool{clusterProfiler}'s GSEA, we also observe the just-described set of analytical choices to trigger a decrease in the adjusted $p$-value of the respective gene sets in the majority of permutations. Note that we additionally observe the approach to the removal of duplicated gene IDs to be a common trigger of decrease. However, the corresponding effect is always negligible. The strongest decreases using \tool{clusterProfiler}'s GSEA are observed in one permutation for both gene sets, each. They lead from $0.47$ to $0.077$ and from $1$ to $0.058$, respectively, such that the corresponding `tweaked' adjusted $p$-values remain just above the significance threshold. \\
\textit{\textbf{True sample labels:}}
Similar to goal~1, a reduction of the adjusted $p$-value cannot be achieved for \tool{GOSeq}, \tool{DAVID}, and \tool{clusterProfiler}'s ORA for any of the gene sets. In contrast, for \tool{GSEAPreranked}, an initially non-significant $q$-value of $0.29$ is reduced to the significant value of $0.20$ through the change of the DE method in data preprocessing. For the web-based application \tool{GSEA}, the initial $q$-value of one of the considered gene sets is $0.13$, indicating significant differential enrichment even before the exploitation of uncertainty. The modification of the weighting pattern of the genes in the computation of the enrichment score leads to a further decrease to a $q$-value of $0.001$.

\subsection{Results for goal~3: minimise rank of a specific gene set}

In our study, we define the rank of a gene set in the GSA results based on the listing of all gene sets which typically stems from the order of their adjusted $p$-values. Thereby, the gene set with the lowest adjusted $p$-value (indicating the strongest association with the condition of interest) occupies the first position. If several gene sets have an identical adjusted $p$-value, we assign them the same rank. Furthermore, we take into account that the number of contained gene sets often varies between GSA results tables, particularly when generated using different GSA methods that often refer to distinct versions of the corresponding gene set database. To ensure comparability between the rank of a gene set across GSA results tables, we therefore divide each rank in the given GSA results table by the maximum assigned rank from that table. The resulting \textbf{relative} ranks range from 0 to 1. Thereby, a lower relative rank indicates that the corresponding gene set’s adjusted $p$-value is generally lower compared to the remaining genes (and vice versa). In particular, an adjusted $p$-value of $1$ automatically results in a relative rank of $1$. In the following, we use the expressions `rank' and `relative rank' interchangeably. \\
\begin{figure}
    \centering    \includegraphics[scale = 0.58]{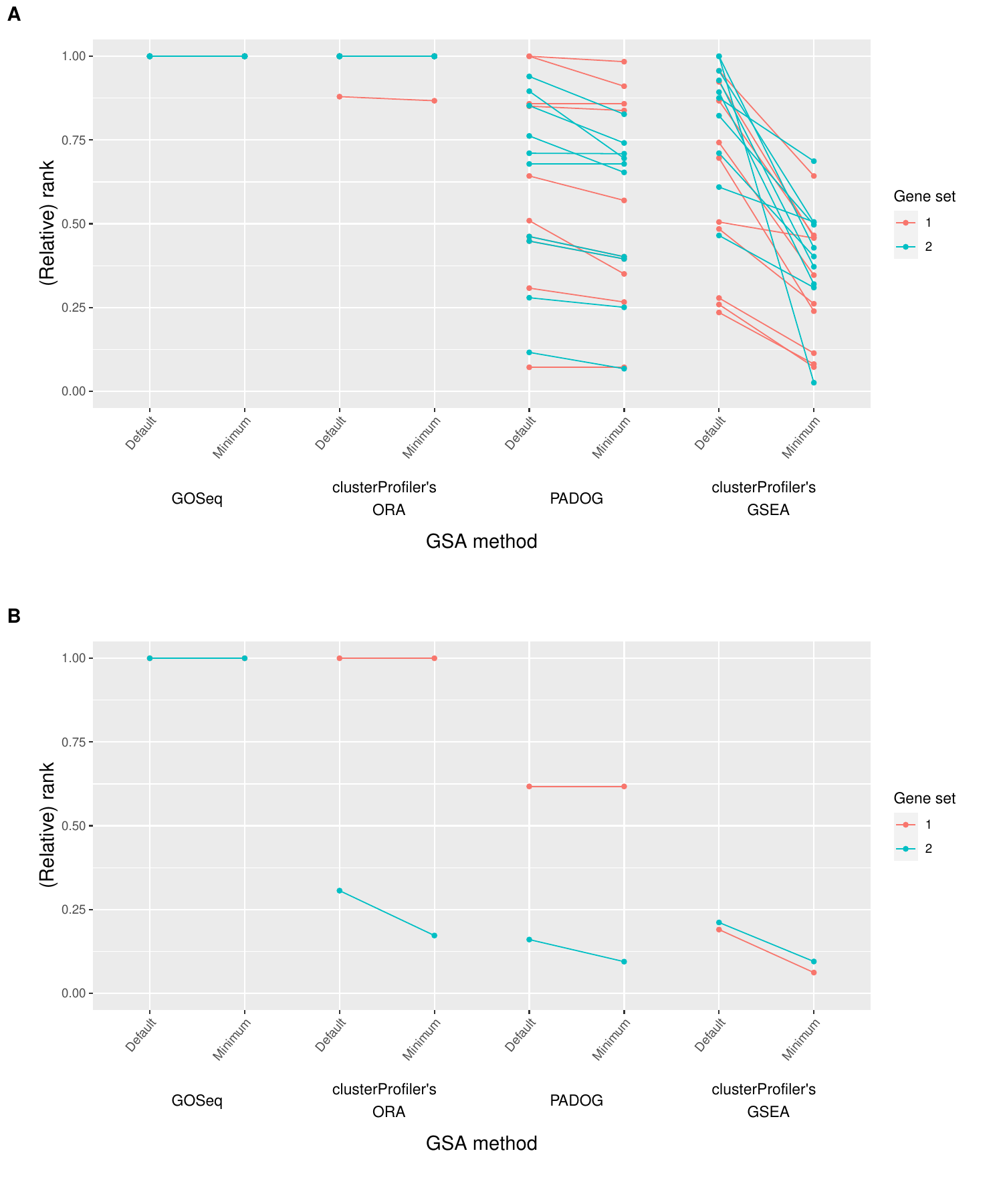}
    \caption{\small Goal~3: The (relative) ranks in the Pickrell data set (`Minimum'), obtained through the exploitation of uncertainty, are compared to the corresponding values resulting from the default analytical choices (`Default'). For each optimisation process, the associated optimised and the default rank are connected through a line. (A) presents the results for the ten random permutations and (B) for the true sample labels. On the $x$-axis, the individual methods investigated in the context of goal~3 are displayed. The results for gene set 1 are shown in red and those for gene set 2 in blue.}
    \label{fig_results_relrank}
\end{figure} 
\textit{\textbf{Random sample label permutations:}} \\
For \tool{GOSeq} and \tool{clusterProfiler}'s ORA, we observe the same pattern as with goals 1 and 2, namely that the ranks of the two considered gene sets remain at the highest possible value of $1$ before and after the exploitation of uncertainties. \\
In contrast, for \tool{PADOG}, we observe many slight to moderate decreases in the relative ranks for the majority of permutations, triggered through the choice of the pre-filtering approach or the method to transform the RNA-Seq data. Note that the relative ranks are generally low even before exploiting any uncertainty, especially compared to the adjusted $p$-values from goal~2; see Figure~\ref{fig_results_padj}. For instance, with all uncertain choices in their default, gene set 2 has a relative rank of $0.12$ in one permutation (whereas the corresponding adjusted $p$-value amounts to $0.82$). This relative rank is further decreased to $0.07$ through the choice of the method to transform the RNA-Seq data, whereas the adjusted $p$-value increases to $0.92$. This indicates that, while the initial adjusted $p$-values of all gene sets are already generally high, the adjusted $p$-values of the majority of the remaining gene sets increase even more strongly through the exploitation of uncertainties than the adjusted $p$-value of gene set 2.\\
We observe notable decreases in the relative ranks across the permutations for \tool{clusterProfiler}'s GSEA. Analogous to optimisation goals 1 and 2, many of these decreases are triggered through the choice of the DE method to generate the required input and the specification of an equal weighting for all genes in the computation of the enrichment score. The observations coincide with the corresponding ones from goal~2, meaning that through the exploitation of the uncertainty, a user does not trigger a decrease in the adjusted $p$-value of all gene sets, but specifically modifies the ranking of the gene sets in the GSA results in favour of their `preferred' gene set. For instance, for gene set 2, there is one permutation in which the choice of the DE method and the specification of an equal weight of all genes in the computation of the enrichment score results in a decrease of the relative rank from $1$ to $0.026$. At the same time, the adjusted $p$-value decreases from $1$ to $0.058$. \\
\textit{\textbf{True sample labels:}}
For \tool{GOSeq}, the relative ranks cannot be decreased from their respective initial value of $1$ for either of the gene sets. In contrast, for \tool{clusterProfiler}'s ORA, a reduction from $0.31$ to $0.17$ is triggered through the choice of the universe for gene set 2. For \tool{PADOG}, the already low relative rank of $0.16$ of gene set 2 can be further decreased to $0.09$ through the choice of the method for RNA-Seq transformation. Note, however, that the corresponding adjusted $p$-values before and after the exploitation of the uncertainties, respectively, amount to $0.92$ and $0.90$. This indicates that the adjusted $p$-values of the vast majority of remaining gene sets are even closer to $1$. \\

\section{Discussion}

In our analysis, we quantified over-optimism effects resulting from the multiplicity of analysis strategies combined with selective reporting for several popular gene set analysis methods. We thereby considered several types of expectancies/hopes the researchers might have when tweaking their analyses, which we translated into three distinct goals. The maximisation of the number of differentially enriched gene sets (goal~1) grants the researcher greater flexibility in generating hypotheses. By minimising a specific gene set's adjusted $p$-value or rank among the remaining gene sets, a researcher attempts to fulfill their expectations as to what constitutes interesting results. \\
Our study shows that the potential for the generation of over-optimistic results in the context of all three goals particularly affects two methods, namely GSEA provided by the \tool{R} package \tool{clusterProfiler} and the web-based method \tool{GSEAPreranked}. Both methods apply variations of Gene Set Enrichment Analysis \citep{subramanian2005gene} commonly known to produce inflated false discovery rates (FDRs) \citep{wu2012camera, maleki2020gene}. However, we observe their frequent use; see e.g. \citet{reimand2019pathway, lopes2020sex}. Our study shows that, in addition to the above-described problem, both methods commonly grant the user to further tweak the results, e.g., by changing the weighting pattern of the genes in the assessment of differential enrichment, a modification easily justifiable through information from corresponding user manuals. Since the evidence from our study casts even further doubt on the reliability of thus made research findings, our study should therefore be considered a reinforcement of the recommendation against these methods. \\
While our work focuses on the exploitation of uncertainty in GSA at the analysis stage, a researcher is in reality already confronted with additional uncertainties during the generation of the gene expression data set. These include (but are not limited to) the choice of the library preparation kit, the quality filter to remove sequences likely to contain errors, and the source of references for gene annotation. A modification in one or more of these aspects can lead to additional variability in the GSA results even when the downstream analysis strategy remains the same. Our focus on later stages of the research process can therefore result in an underestimation of the potential for over-optimistic results. The other way around, the strictly systematic manner of optimising the results as employed in our work does certainly not fully reflect the reality of data analysis practice, which in turn suggests that our study might overestimate over-optimism. \\ 
Firstly, the probability of exploiting an existing uncertainty is likely to vary between the different steps, depending on the required effort as well as the researcher's programming expertise and mindset. For example, the adjustment of a preprocessing step, such as changing the method for differential expression analysis when generating the required input for ORA (which requires modifying several lines of code), generally comes with a greater effort compared to changing a parameter value, which can often be done by a single click. Furthermore, the needed efforts are generally higher if the researcher is equipped with little programming experience. The amount of tweaking a real researcher may realistically perform in a practical project is therefore individual. Our selection of analysis steps and options the fictive researcher may choose from does not take this individual variability into account and thus inevitably involves some arbitrariness. This arbitrariness extends to other aspects such as the specification of default options, despite being based on extensive literature research. Most importantly, it may be argued that researchers would not consider as many options in practice. \\
Secondly, it is more likely that a real researcher adopts a change in a flexible step if it leads to a noticeable improvement in the results instead of (as is the case in our study design) accepting each improvement, however small it may be. Indeed, researchers know that each change from the default might require a justification eventually, and may thus be reluctant to engage in changes that do not bring substantial benefits. \\
Thirdly, we consider the three optimisation goals separately, while a real researcher might have several of these goals in mind. For instance, they might want to increase the number of differentially enriched gene sets while, at the same time, trying to reduce the adjusted $p$-value of a specific gene set. Likewise, they may have two or more gene sets in mind whose relevance they want to simultaneously increase in the GSA results. On the one hand, considering a single optimisation goal as fixed, as in the fashion of our study, might lead to an underestimation of the potential for over-optimistic results since we ignore an additional source of multiplicity. However, pursuing multiple goals simultaneously might make it more difficult to induce satisfactory results in a cherry-picking manner, such that the resulting over-optimism in the GSA results may actually be less pronounced. In future work, it would therefore be interesting to investigate the effect on the level of over-optimism resulting from the consideration of multiple optimisation goals. \\
Fourthly, in the context of goal~1, there are several optimisation processes in which the {\it default} number of differentially enriched gene sets (i.e. before performing any optimisation) is already substantial. It could be increased even further for many of these cases; in reality, however, a researcher would be unlikely to increase an already high number of significant results. While a higher number of differentially enriched gene sets offers a higher flexibility regarding the storyline of the paper and biological understanding of the results, it can also complicate their interpretation and reporting, even forcing the researcher to report only a subset of the detected gene sets. Our study does not take this into account and assumes that the researcher would always be interested in increasing the number of significant gene sets. \\
To sum up, our study unavoidably required a certain amount of simplifications, which may not only imply an underestimation of the over-optimistic effect of interest but also its overestimation.   \\
A further aspect of our study that is subject to arbitrariness is the choice of the gene sets whose adjusted $p$-value or rank (for optimisation goals 2 and 3, respectively) we attempted to minimise. Different choices of gene sets might lead to different extents to which over-optimistic results can be achieved and the gene sets considered in our study only make up a small fraction of the gene sets provided by the gene set database in the individual methods. Furthermore, for goals 2 and 3, our study focused mainly on gene sets provided by the gene set database GO (with subontology `Molecular Function'). It would therefore be interesting to extend the selection of gene sets, additionally considering gene sets from other gene set databases. \\
To address the arbitrariness in our study design when \textit{mimicking a hypothetical researcher}, it would be interesting to perform a real-life multianalyst experiment in the spirit of \citet{silberzahn2018many}. This would imply recruiting several teams of analysts and presenting them with the same research question and RNA-Seq data set. The task would then be to investigate the variation in the different steps of the chosen analysis strategies and in the results between the teams. This experiment would additionally allow for an assessment of the meaningfulness of the tweaked GSA results in the context of the conditions of interest. Note that such experiments may also be conducted with students as analysts in the context of undergraduate teaching, see for example \citet{heyman2022multiverse}. \\
Despite its limitations, our study clearly encourages readers of publications presenting GSA results to interpret them cautiously---i.e. with the inherent uncertainties and possible over-optimism in mind. Our study reveals that for some methods, it is relatively easy to cherry-pick in the context of GSA and that the resulting over-optimism is sometimes substantial. Although we obviously cannot provide evidence of the extent of cherry-picking in practice (since it happens behind closed doors and is naturally never reported), we conjecture that its incidence is non-negligible. \\ 
What can `real' researchers do to avoid over-optimism when conducting gene set analysis? 
It is important to not only report those analyses that are identified as `most preferable' \textit{after} inspecting the results. GSA is often performed as part of exploratory research, which makes it difficult to decide on all aspects and details of the analysis before running it. Helpful guidance---may it be in the form of neutral comparison studies supporting the method's choice or user manuals provided alongside the GSA methods---is often scarce, making the prespecification of the analysis strategy even more complex. In this context, it is certainly not realistic to require researchers to fix all analytical choices in advance. A valid alternative approach would be to report the GSA results obtained with several analysis strategies in an effort to transparently disclose and integrate the underlying uncertainties while refraining from reporting only the `best' ones in a cherry-picking manner. Alternatively, one may select the analysis strategy after running the analyses---and thus identifying potential problematic behaviours of some of the analysis strategies---but \textit{without} looking at the `main results' (which is formalised in terms of `goals' in our study). This latter approach can be seen as a relatively safe compromise between the full prespecification of the analysis strategy, which lacks flexibility in an exploratory setting, and the results-driven selection of analysis strategies, which often leads to over-optimism as demonstrated through our study. \\

\section*{Funding Information}

{\label{974317}}
This work is supported in part by funds from the German Research Foundation (DFG: BO3139/7-1 and BO3139/9-1).

\section*{Acknowledgements}

{\label{749861}}

The authors thank Savanna Ratky for language correction.

\section*{Competing interests}
No competing interest is declared.

\section*{Data availability statement}
The data that support the findings of this study are openly available on \href{https://github.com/chillemille/OverOptimism_in_GeneSetAnalysis}{\color{bleudefrance}{GitHub}}.

\FloatBarrier
\bibliographystyle{apalike}
\bibliography{reference}

\newpage
\beginsupplement
\textbf{\Large Supplemental Material}

\section{Additional sources of uncertainty}

As mentioned in the main document, there are two additional sources of uncertainty in the framework of \citet{hoffmann2021multiplicity} a researcher faces when carrying out GSA. However, they occur in the process of \textit{generating} the gene expression data (which we assume to be complete so that we do not include these uncertainties in our study). 

\textit{Measurement uncertainty}, on the one hand, refers to the circumstance that the measurements contained by the data set are affected by a certain amount of imprecision.  \textit{Sampling uncertainty}, on the other hand, emerges since the data set is assumed to be a random sample of the population of interest, inducing a certain variability. \\

\section{Exploited uncertainty}
In the following, we elaborate on the analytical choices affected by uncertainty in data preprocessing (to format the available gene expression data set as required by the individual GSA methods) and the parameters that we exploit in our study. The options specified as the default to the corresponding step in our study are underlined. For a more detailed description of all of these aspects, see our earlier work \citep{wuensch2023rna}. 

\subsection{Exploited data preprocessing uncertainty}\label{sec_exploited_datapreprocess}

For each of the uncertain choices in data preprocessing, the set of options was acquired through an extensive review of commonly used methods and approaches in published scientific literature. If for some steps, corresponding literature is scarce or even non-existent, we extended our review to online communities. An overview of the exploited data preprocessing uncertainties per GSA method is provided in Table \ref{uncertainties_preprocess}.\\

\textit{\textbf{Pre-filtering}} \\
Pre-filtering refers to the exclusion of lowly expressed genes, i.e. genes with a low magnitude of count data. One of several reasons to pre-filter a gene expression data set is that lowly expressed genes are unlikely to be detected as differentially expressed. Pre-filtering approaches are particularly suggested alongside DE methods and often differ between them. For DESeq2, on the one hand, a manual approach is proposed in which all genes with less than a pre-specified number of read counts are excluded from further analysis. For limma (and edgeR), pre-filtering is typically performed using the function \texttt{filterByExpr()}, which excludes all genes from further analysis with less than a pre-specified counts-per-million in a pre-specified number of samples. For those GSA methods that are not preceded by a DE method, i.e. the web-based application \tool{GSEA} as well as \tool{PADOG}, the two just-described approaches are selected as the two options: 
\begin{itemize}[noitemsep]
    \item[(i)] \underline{option 1}: remove all genes with less than ten read counts across all samples, 
    \item[(ii)] option 2: pre-filtering using edgeR's function \texttt{filterByExpr()}. 
    \end{itemize}
For the remaining GSA methods in our study, the options for pre-filtering, therefore, depend on the optimal DE method in the specific optimisation process, such that the options for pre-filtering in the case of DESeq2 being the optimal DE method are 
\begin{itemize}[noitemsep]
    \item[(i)] \underline{option 1}: remove all genes with less than ten read counts across all samples, 
    \item[(ii)] option 2: remove all genes with less than 50 read counts across all samples. 
    \end{itemize}
If, on the other hand, limma is the optimal DE method for an optimisation process, then the options to pre-filtering are 
\begin{itemize}[noitemsep]
 \item[(i)] \underline{option 1}: pre-filtering using edgeR's function \texttt{filterByExpr()}, 
    \item[(ii)] option 2: remove all genes which do not have at least $1$ count per million in at least two samples. 
\end{itemize}
Note that option 2 stems from an older user guide provided by edgeR in 2018. \\

\textit{\textbf{Removal of duplicated gene IDs}}\\
There are a multitude of formats to identify the genes in the gene expression data set and if the format in the given data set differs from the format(s) accepted by the chosen GSA method, a conversion to the required format is necessary. This conversion results in duplications of some gene IDs which must be removed manually by the user. The two utilised options for removing these duplicates are based on our earlier work \citep{wuensch2023rna} and correspond to

\begin{itemize}[noitemsep]
 \item[(i)] \underline{option 1}: keep the gene ID occurring first, 
\item[(ii)] option 2: in each sample, keep the rounded mean expression value of all genes associated with the duplicated gene ID.
\end{itemize}

\textit{\textbf{Differential expression analysis}} \\
Differential expression analysis is typically performed for 
ORA methods to obtain the required input in the form of a list of differentially expressed genes. However, it is also commonly used for FCS methods which require as input a ranking of all genes from the experiment based on their magnitudes of differential expression. In our study, we utilise two popular (and parametric) DE methods, namely 
\begin{itemize}[noitemsep]
    \item[(i)] \underline{option 1}: DESeq2 \citep{love2014moderated},
    \item[(ii)] option 2: voom/limma \citep{law2014voom}.
\end{itemize}
As described above, we specify the options to pre-filtering in a specific optimisation process depending on the optimal DE method.  \\

\textbf{\textit{Transformation (and normalisation)}} \\
Those FCS methods investigated in our study that require as input the gene expression data set as a whole were initially developed for microarray measurements. Therefore, they assume different characteristics than those that are rooted in RNA-Seq measurements. A common way to handle this discrepancy is to transform the RNA-Seq data such that the characteristics of the transformed data (approximately) match those of microarray measurements. We have selected two methods for RNA-Seq transformation based on our earlier work \citep{wuensch2023rna}, namely 
\begin{itemize}[noitemsep]
    \item[(i)] \underline{option 1}: voom transformation as part of the work of \citet{law2014voom},
    \item[(ii)] option 2: variance stabilising transformation as proposed by \citet{love2014moderated}. 
\end{itemize}
Note that both transformation methods additionally normalise the RNA-Seq measurements. For a more detailed description of both transformation methods, see our earlier work \citep{wuensch2023rna}.

\begin{table}
\caption{Overview of the exploited data preprocessing steps per computational GSA method. Note that we use the term `\tool{cP}' as an abbreviation for the \texttt{R} package `\tool{clusterProfiler}'. As described in the main document, we have reduced the number of exploited preprocessing steps for the web-based applications \tool{DAVID}, \tool{GSEA}, and \tool{GSEAPreranked} since the corresponding optimisation processes are performed by hand.} \label{uncertainties_preprocess}
\small \centering

   \renewcommand{\arraystretch}{1} 
    \begin{tabularx}{0.74\textwidth}{p{0.15\linewidth}Y{0.03\linewidth}Y{0.03\linewidth} Y{0.03\linewidth}Y{0.03\linewidth}Y{0.03\linewidth}Y{0.01\linewidth}Y{0.03\linewidth}Y{0.045\linewidth}Y{0.03\linewidth}Y{0.045\linewidth}}

    \toprule
   \multicolumn{1}{c}{} & & & \multicolumn{3}{c}{ORA methods}&  & \multicolumn{4}{c}{FCS methods}\\

     \addlinespace[0.2cm]
     \cline{4-6} \cline{8-11}
     \addlinespace[0.2cm]

Preprocessing step &   & & \begin{turn}{45}\tool{DAVID} \end{turn}& \begin{turn}{45}\tool{GOSeq} \end{turn}& \begin{turn}{45} \tool{cP}'s ORA \end{turn} & &\begin{turn}{45}\tool{GSEA} \end{turn} & \begin{turn}{45} \tool{GSEAPreranked} \end{turn}&\begin{turn}{45} \tool{PADOG} \end{turn} & \begin{turn}{45} \tool{cP}'s GSEA \end{turn} \\
    \midrule

\rowcolor{gray!25}    
Pre-filtering  & & &  & \newcrossmark & \newcrossmark & & \newcrossmark &  & \newcrossmark & \newcrossmark \\
\rowcolor{gray!7}
Removal of \newline duplicated gene IDs &  & &  &  & \newcrossmark &  &  &  & \newcrossmark & \newcrossmark \\
\rowcolor{gray!25}
Differential expression analysis  & & & \newcrossmark & \newcrossmark &  \newcrossmark & &  & \newcrossmark &  & \newcrossmark  \\
\rowcolor{gray!7}
Transformation \newline (and normalisation) & & &  &  &  & & \newcrossmark &  & \newcrossmark  &  \\

 \bottomrule
    \end{tabularx}
    \end{table}

\subsection{Exploited parameter uncertainty}
 
In the following, we elaborate on the five parameter choices affected by uncertainty that we exploit in our study. For an additional overview, see Table \ref{parameter_uncertainties}. While some parameters are flexible across ORA and FCS, others are specific to one of the two general approaches. In particular, no parameter can be chosen flexibly for all considered GSA methods. For all of these parameters apart from the gene set database, a default value is proposed by the corresponding method. A detailed description of these parameters can be found in our earlier work \citep{wuensch2023rna}. Note that for each uncertain parameter, we choose the number of options within a range that we regard as realistic considering that a (well-intentioned) researcher is unlikely to try indefinitely many options. \\

\textbf{\textit{Geneset database}} \\
In this work, we focus on the choice between the following two commonly used gene set databases which are offered by (almost) all of the considered  GSA methods:
\begin{itemize}[noitemsep]
    \item[(i)] \underline{option 1}: Gene Ontology (`GO') with subontology `Molecular Function' \citep{ashburner2000gene, gene2023gene}, 
    \item[(ii)] option 2: KEGG \citep{kanehisa2000kegg, kanehisa2023kegg}. 
\end{itemize}
We do not make use of the opportunity to upload a user-defined gene set database which is offered by some GSA methods. Note that the individual GSA methods do not automatically refer to identical versions of the respective gene set database since the methods are not maintained in identical time intervals. As a consequence, the number of gene sets (and therefore also the set of gene sets) contained by a specific gene set database can differ between the considered GSA methods. \\

\textit{\textbf{Universe}} \\
The flexible choice of the universe (also called `background') is specific to the GSA methods classified as ORA since typically only a subset of the genes from the experiment are provided as input for the corresponding methods. This leads to a loss of information on the entirety (i.e. `universe') of the genes. However, in ORA, this information is typically required as the `population' in the hypergeometric distribution which serves as the underlying null distribution.  The information on the universe must therefore be obtained alternatively. In our study, we utilise the following options as the universe: 
\begin{itemize}[noitemsep]
    \item[(i)] \underline{option 1}: all genes that are part of the chosen gene set database, 
    \item[(ii)] option 2: all genes from the initial gene expression data that are tested for differential expression. 
\end{itemize}

\tool{GOSeq} constitutes an exception to the `regular' ORA methods as its required input corresponds to all genes from the experiment labelled by their status of differential expression. This means that the information on the entirety of the genes is, in contrast to the `regular' ORA methods, available. The options of the universe utilised in our study for \tool{GOSeq} are therefore the following: 
\begin{itemize}[noitemsep]
    \item[(i)] \underline{option 1}: all genes from the input that are members to at least one gene set from the chosen gene set database, 
    \item[(ii)] option 2: all genes from the input.  
\end{itemize}

\textbf{\textit{Method}} \\
\tool{GOSeq} is the only method among the selection in which the user can modify the method to obtain the $p$-values of enrichment of the gene sets. In our study, we exploit the following options for \tool{GOSeq}: 
\begin{itemize}[noitemsep]
    \item[(i)] \underline{option 1}: Wallenius distribution \citep{wallenius1963biased},
    \item[(ii)] option 2: resampling method.
\end{itemize}
The Wallenius distribution is an approximative approach and an extension to the hypergeometric distribution while the alternative resampling method computes the $p$-values of enrichment in a manner similar to FCS methods. \\

\textbf{\textit{Gene-level statistic}} \\
For FCS methods, the gene-level statistic is the metric (i.e. `formula') used to generate the ranking of the genes based on each of their magnitude of differential expression between the conditions. For those FCS methods that generate the ranking internally (\tool{GSEA} and \tool{PADOG} in our study), it is generated directly from the initial gene expression data set. For the remaining FCS methods which require the user to create the ranking externally as part of data preprocessing (\tool{GSEAPreranked} and \tool{clusterProfiler}'s GSEA), it is typically created from the results of differential expression analysis. For the former, the choice of a gene-level statistic often presents a flexible parameter within the method, contributing to parameter uncertainty. The uncertain choice of a gene-level statistic for the latter, on the other hand, indicates data preprocessing uncertainty as opposed to parameter uncertainty (note however, in our study, we use a fixed gene-level statistic for these FCS methods in the data preprocessing). \\
Note that for \tool{PADOG}, the gene-level statistic does not present any flexibility. Therefore, the web-based application \tool{GSEA} is the only FCS method offering flexibility in the parameter `gene-level statistic'. The corresponding options are 
\begin{itemize}[noitemsep]
    \item[(i)] \underline{option 1}: signal-to-noise ratio,
    \item[(ii)] option 2: t-statistic,
    \item[(iii)] option 3: difference of classes.
\end{itemize} 

\textbf{\textit{Weight}} \\
The parameter `weight' is related to the strength of the contribution of each gene in the assessment of differential enrichment of a given gene set. For \tool{GOSeq}, the weight of each gene enters the analysis through a probability weighting function. In our studies, we utilise the following options for \tool{GOSeq}
\begin{itemize}[noitemsep]
    \item[(i)] \underline{option 1}: assign a higher weight to genes with shorter transcript length, 
    \item[(ii)] option 2: assign a higher weight to the genes with a lower gene expression level in the RNA-Seq gene expression data set. 
\end{itemize}
Option 1 accounts for the circumstance that genes with longer transcript length are more likely to be detected as differentially expressed, leading to a higher likelihood of detection of differential enrichment of those gene sets whose members are generally longer in transcript length. On the other hand, option 2 accounts for all possible biases leading to a decreased statistical power in detecting the gene sets as differentially enriched that contain many genes with a low overall expression level. \\
In the context of the FCS methods, the weight of the genes can be modified for those methods that are based on the method Gene Set Enrichment Analysis (\tool{GSEA}, \tool{GSEAPreranked}, and \tool{clusterProfiler}'s GSEA), namely by adapting the exponent in the calculation of the enrichment score (see \citet{subramanian2005gene} or \citet{wuensch2023rna} for a more detailed description). In our study, we utilise as options for the exponent value
\begin{itemize}[noitemsep]
    \item[(i)] \underline{option 1}: exponent 1. Weight each gene by its absolute value of the gene-level statistic.
    \item[(ii)] option 2: exponent 0. Assign each gene the same weight.
    \item[(iii)] option 3: exponent 1.5. Weight each gene by the $1.5$-times exponentiated value of the gene-level statistic. 
    \item[(iv)] option 4: exponent 2. Weight each gene by the squared value of the gene-level statistic.
\end{itemize}

\begin{table}

\caption{Overview of the exploited parameters per computational GSA method. Note that `\tool{cP}' is the abbreviation for the \texttt{R} package \tool{clusterProfiler}. The asterisk `$\ast$' indicates that the choice of gene set database is exploited for optimisation goal 1 only.}\label{parameter_uncertainties}
\small \centering

   \renewcommand{\arraystretch}{1} 
    \begin{tabularx}{0.78\textwidth}{p{0.2\linewidth}Y{0.03\linewidth}Y{0.03\linewidth} Y{0.03\linewidth}Y{0.03\linewidth}Y{0.04\linewidth}Y{0.0001\linewidth}Y{0.04\linewidth}Y{0.045\linewidth}Y{0.03\linewidth}Y{0.045\linewidth}}

    \toprule
   \multicolumn{1}{c}{} & & & \multicolumn{3}{c}{ORA methods}&  & \multicolumn{4}{c}{FCS methods}\\

     \addlinespace[0.2cm]
     \cline{4-6} \cline{8-11}
     \addlinespace[0.2cm]

Preprocessing step &   & & \begin{turn}{45}\tool{DAVID} \end{turn}& \begin{turn}{45}\tool{GOSeq} \end{turn}& \begin{turn}{45} \tool{cP}'s ORA \end{turn} & &\begin{turn}{45}\tool{GSEA} \end{turn} & \begin{turn}{45} \tool{GSEAPreranked} \end{turn}&\begin{turn}{45} \tool{PADOG} \end{turn} & \begin{turn}{45} \tool{cP}'s GSEA \end{turn} \\
    \midrule

\rowcolor{gray!25}
Gene set database*  & & & \newcrossmark & \newcrossmark & \newcrossmark & & \newcrossmark & \newcrossmark &  & \newcrossmark \\
\rowcolor{gray!7}
Universe &  & & \newcrossmark  & \newcrossmark  & \newcrossmark  &  &  &  &  &  \\
\rowcolor{gray!25}
Method  & & & & \newcrossmark &   & &  &  &  &   \\
\rowcolor{gray!7}
Gene-level statistic & & &  &  &  &&\newcrossmark &   & &     \\
\rowcolor{gray!25}
Weight  & & & & \newcrossmark &  & & \newcrossmark & \newcrossmark &  & \newcrossmark \\

 \bottomrule
    \end{tabularx}
    \end{table}

\section{Design of the study}

\subsection{Settings}

\textit{\textbf{Gene expression data sets}}\\
For our study, we have selected two RNA-Seq data sets based on the following criteria: 
\begin{itemize}[noitemsep]
    \item[(i)] the conditions of the samples are known and binary,
    \item[(ii)] there are at least ten samples per condition,
    \item[(iii)] the sample sizes per condition are similar,
    \item [(iv)] the gene expression measurements stem from an organism supported by each of the investigated computational methods. 
\end{itemize}
Criteria (ii) and (iii) ensure that we can generate a sufficient number (i.e. ten) of random permutations of the true sample labels whose assignments to the individual samples differ from the true assignments in sufficiently many cases. \\
\textbf{\textit{Gene sets (for goals 2 and 3)}}\\
For illustration, we initially intended to select gene sets for the Pickrell and Bottomly data set, respectively, from published application papers that are similar in terms of the underlying biological context and research question. As the match for the Pickrell data set, we chose the work of \citet{lopes2020sex} who use \tool{GSEAPreranked} to identify sex-biased enrichment in $29$ tissue types ($28$ solid tissues, including adipose (subcutaneous), thyroid, and whole blood), based on $8279$ tissue samples. The data set contains gene expression measurements of $n=188$ female and $n=360$ male subjects and the expression profiles of $30243$ genes. As the gene sets whose adjusted $p$-value and rank we attempt to optimise for goals 2 and 3, we chose two gene sets with the highest magnitude of differential enrichment between males and females in the tissue `whole blood'. This selection resulted in the gene sets  \textit{Demethylation} and \textit{T cell mediated immunity}, both from gene set database GO with subontology `Biological Process'. \\
For the RNA-Seq data set provided by \citet{bottomly2011evaluating}, we 
selected the work of \citet{kraus2012mouse} as the corresponding match. Their work includes three Panther gene ontology (GO) analyses that compare the gene expression level between entire mouse embryos of four strains, eviscerated mouse embryos of four strains, and eviscerated embryos of eleven strains. In their work, gene expression is measured in the form of microarray measurements. Note that mouse strain `C57BL/6J' is among the considered mouse strains in all three analyses, whereas strain `DBA/2J' is not included. We have selected the two gene sets with the highest relevance in all three results, namely \textit{Metabolic process} and \textit{Cellular Process}. Both gene sets are provided by the gene set database Gene Ontology with subontology Biological Process.  \\
However, several issues in the attempt to minimise the adjusted $p$-values and ranks of these gene sets selected for the Pickrell and Bottomly data set force us to resort to alternative gene sets in these cases. For an overview of the selected gene sets for each GSA method and gene expression data set, see Table \ref{GeneSet_selection}.\\
Firstly, when providing a gene expression data set stemming from the mouse organism, \tool{GSEA} and \tool{GSEAPreranked} internally convert the mouse gene IDs to orthologous human gene IDs. Differential enrichment is, therefore, assessed for `human' gene sets. Consequently, we cannot work with either of the `mouse' gene sets extracted from \citet{kraus2012mouse} when working with the Bottomly data set. In these cases, we, therefore, use the corresponding `human' gene sets from \citet{lopes2020sex}, instead. \\
For \tool{clusterProfiler}'s GSEA, neither of the two gene sets from \citet{kraus2012mouse} are included by GO, forcing us to proceed in the same manner as with \tool{GSEA} and \tool{GSEAPreranked} for the Bottomly data set. For \tool{clusterProfiler}'s ORA, both gene sets from \citet{kraus2012mouse} are also not included in any of the results tables. However, we cannot assess whether both gene sets are indeed not even provided by the gene set database GO since the results table provided by \tool{clusterProfiler}'s ORA generally only include those gene sets containing at least one differentially expressed gene set from the input. However, because both gene sets are not included by GO when working with \tool{clusterProfiler}'s GSEA, we also resort to both gene sets matching the Pickrell data set for \tool{clusterProfiler}'s ORA. 
Finally, \tool{PADOG} does not offer the gene set database GO internally so we have to resort to KEGG, leading to the choice of the gene sets `Primary immunodeficiency' and `Graft versus host disease'. Both gene sets are among the most differentially enriched KEGG pathways in the tissue `whole blood'. Since when using \tool{PADOG}, the considered gene sets are the same for gene expression data stemming from human and mouse, we use these gene sets for both gene expression data sets. 

\begin{table}[ht]
\small
\centering
    \caption{The gene sets whose adjusted $p$-values and ranks are optimised for goals 2 and 3, respectively. The representation is clustered by the GSA methods for which the same gene sets are optimised.} \label{GeneSet_selection}
    \begin{tabularx}{0.99\textwidth}{p{0.3\textwidth}p{0.3\textwidth}p{0.3\textwidth}}
    \toprule
      GSA method & Gene sets (1 and 2) \newline optimised \newline in Pickrell data set & Gene sets (1 and 2) \newline optimised \newline in Bottomly data set \\
         \toprule
         \rowcolor{gray!25}
         \tool{GOSeq}  & T cell mediated immunity \newline Demethylation & Cellular process \newline Metabolic process \\
         \rowcolor{gray!7}
       \tool{DAVID} & T cell mediated immunity \newline Demethylation & Cellular process \newline Metabolic process \\
        \rowcolor{gray!25}
          \tool{clusterProfiler}'s ORA & T cell mediated immunity \newline Demethylation & T cell mediated immunity \newline Demethylation\\
       \rowcolor{gray!7}
       \tool{PADOG} & Primary immunodeficiency \newline Graft versus host disease & Primary immunodeficiency \newline Graft versus host disease \\ 
       \rowcolor{gray!25}
        \tool{clusterProfiler}'s GSEA & T cell mediated immunity \newline Demethylation & T cell mediated immunity \newline Demethylation \\
       \rowcolor{gray!7}
       \tool{GSEA} & T cell mediated immunity \newline Demethylation & T cell mediated immunity \newline Demethylation \\
       \rowcolor{gray!25}
       \tool{GSEAPreranked} & T cell mediated immunity \newline Demethylation & T cell mediated immunity \newline Demethylation \\

        \bottomrule   
    \end{tabularx} \\ 

    \label{matched_genesets}
\end{table}

\section{Results}
\subsection{Illustration of one optimisation process}
We describe the progression of one specific optimisation process resulting from the setting consisting of
\begin{itemize}[noitemsep]
    \item[(i)] goal 1: maximise the number of differentially enriched gene sets
    \item[(ii)] gene expression data set: Pickrell data set 
    \item[(iii)] sample labels: random sample label permutation 6
    \item[(iv)] method: \tool{clusterProfiler}'s GSEA.  
\end{itemize}

\begin{figure}[h]
    \centering
    \includegraphics[scale = 0.6]{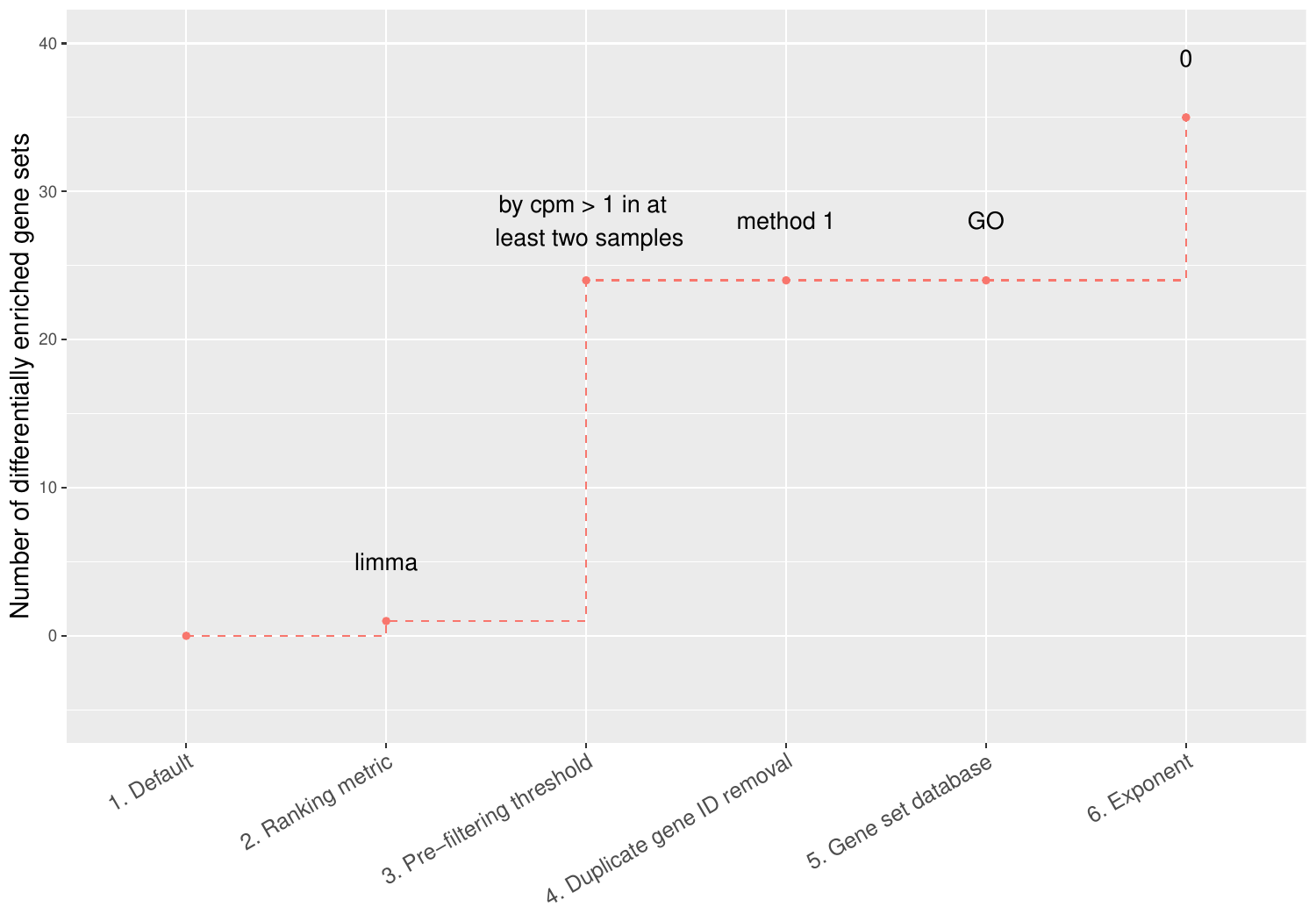}
    \caption{Illustration of the progression of one optimisation process following goal 1 in a step diagram. The individual optimisation steps are displayed on the $x$-axis and the corresponding optimal analytical choices are indicated above the steps. This optimisation process takes place in the setting of optimisation goal 1, the Pickrell data set with the sample labels corresponding to permutation 6, and \tool{clusterProfiler}'s GSEA.}
    \label{fig_step_diagram}
\end{figure}

As displayed in Figure 2 in the main document, we exploit three analytical choices associated with data preprocessing uncertainty and two associated with parameter uncertainty. The first three choices consist of the DE method to generate the input ranking of the genes, the approach to pre-filtering, and the approach to removing the duplications of the gene IDs (resulting from gene ID conversion). On the other hand, we exploit the choice of the gene set database and the exponent in the computation of the enrichment score as part of parameter uncertainty. The corresponding default choices are 
\begin{itemize}[noitemsep]
    \item[(i)] DE method: DESeq2,
    \item[(ii)] pre-filtering approach: remove all genes with less than $50$ read counts across all samples, 
    \item[(iii)] approach to the removal of duplicated gene IDs: keep the corresponding gene ID among the duplicates that occurs first,
    \item[(iv)] gene set database: GO (with subontology Molecular Function),
    \item[(v)] exponent value: 1, i.e. each gene is weighted by their absolute value of the gene-level statistic in the computation of the enrichment score.
\end{itemize}
From Figure~\ref{fig_step_diagram}, in which a graphical illustration of the progression of the optimisation process is provided, we observe that $0$ differentially enriched gene sets result from these default choices.  \\ 
\textbf{\textit{Step 1: optimise the choice of the DE method}} \\
When specifying the alternative method limma as the DE method from which the ranking of the genes is generated, the number of DEGS increases to $1$ compared to when DESeq2 is specified as the DE method. We therefore set \textbf{limma} as the optimal choice of the DE method and update the current optimal results to \textbf{$1$ DEGS}. \\
\textbf{\textit{Step 2: optimise the choice of the pre-filtering approach}} \\
As described in Section \ref{sec_exploited_datapreprocess}, the options to pre-filtering depend on the optimal DE method as determined in the previous step of the optimisation process. The number of DEGS of $1$ at the beginning of this second optimisation step is therefore based on the default pre-filtering approach for limma. This approach corresponds to pre-filtering using the function \textit{filterByExpr()} from the \texttt{R} package \texttt{edgeR} \citep{robinson2010edger} while the alternative option consists of removing all genes which do not exceed $1$
`count-per-million' in at least two samples. This alternative pre-filtering approach results in $24$ DEGS, therefore exceeding $1$ DEGS from the default pre-filtering approach. We, therefore, set \textbf{pre-filtering by cpm $>$  $1$ in at least two samples} as the optimal pre-filtering approach and update the current optimal results to \textbf{$24$ DEGS}. \\
\textbf{\textit{Step 3: optimise the choice of the approach to the removal of duplicated gene IDs}} \\
The alternative approach to the removal of duplicated gene IDs, in which for each duplication the (rounded) mean of read counts of associated genes is kept, cannot exceed the default approach in terms of the number of DEGS. The optimal approach to pre-filtering in this optimisation process therefore corresponds to the default approach of \textbf{keeping the gene ID that occurs first} and the current optimal results remain at \textbf{$24$ DEGS}. \\
\textbf{\textit{Step 4: optimise the choice of the gene set database}}\\
The alternative gene set database KEGG does not lead to an increased number of DEGS compared to the default gene set database GO (with subontology `Molecular Function'). The current optimal results therefore remain at \textbf{$24$} DEGS and the optimal gene set database for this optimisation process is \textbf{GO (with subontology `Molecular Function')}. \\
\textbf{\textit{Step 5: optimise the choice of the exponent in the calculation of the enrichment score}} \\
Assigning each gene the same weight in the computation of the enrichment score (as opposed to weighting each gene by its absolute level of the gene-level statistic) leads to an increase in the number of DEGS to $35$, which is the highest among all alternative options. Consequently, we set \textbf{exponent 0} as the optimal exponent choice and the \textbf{final optimal results} are then \textbf{$35$ DEGS}. \\
\textbf{\textit{Summary}} \\
Through the exploitation of uncertainty in five analytical choices (three choices in data preprocessing and two parameter choices), we can tweak the number of DEGS from $0$ to $35$.

\subsection{Results for the Bottomly data set}

\subsubsection{Results for goal 1: maximise number of differentially enriched gene sets (DEGS)}

\begin{figure}
    \centering    \includegraphics[scale = 0.6]{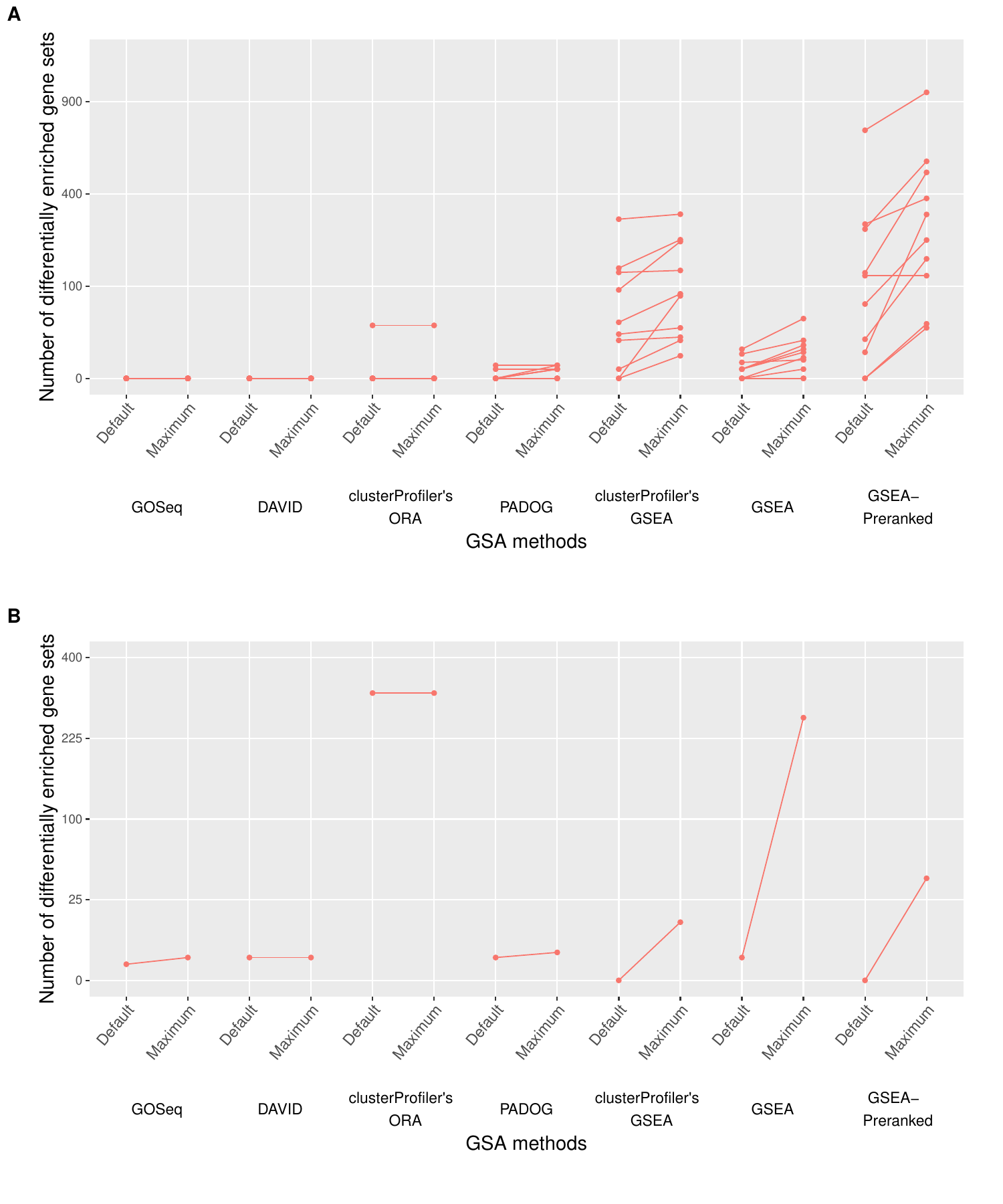}
    \caption{Goal 1: The optimised numbers of differentially enriched gene sets (`Maximum') in the Bottomly data set, obtained through the exploitation of uncertainty, are compared to the corresponding numbers resulting from the default analytical choices (`Default'). For each optimisation process, the associated optimised and the default number are connected through a line. (A) presents the results for the ten random permutations and (B) for true sample labels. On the $x$-axis, the individual methods investigated in the context of goal 1 are displayed. Note the transformation of the scale of the $y$-axis to represent the number of DEGS of different orders of magnitude.}
    \label{fig_results_n_DEGS_supp}
\end{figure}

For an overview of the results, inspect Figure~\ref{fig_results_n_DEGS_supp}. \\
\textit{\textbf{Random sample label permutations:}} 
For \tool{GOSeq} and \tool{clusterProfiler}'s ORA, we observe results similar to the Pickrell data set, namely that no over-optimistic results can be induced in any of the permutations. In particular, the numbers of DEGS amount to $0$ before and after exploiting any uncertainties in the vast majority of permutations. For \tool{PADOG}, our observations also agree with those for the Pickrell data set. An increase from an initial number of $0$ DEGS can be obtained for some of the permutations through the choice of the pre-filtering approach and the method to transform the RNA-Seq data. As with the Pickrell data set, these do not exceed $2$ DEGS after exploitation of uncertainty in any of the permutations. \\
As with the Pickrell data set, the GSEA-based methods, again, indicate a higher potential for over-optimistic results compared to the remaining methods. Among those, it is particularly high for \tool{GSEAPreranked}, for which notable increases are obtained in all but one sample label permutation. The same set of analytical choices triggering increases can be observed as in the Pickrell data set, namely the choice of the DE method and assigning equal weights to all genes in the computation of the enrichment score. This way, a particularly strong increase from $8$ to $316$ DEGS increase can be triggered in one permutation. \\
For \tool{clusterProfiler}'s GSEA and the web-based method \tool{GSEA}, we also observe several notable increases in the number of DEGS. While in absolute terms, the induced increases are generally stronger for the former method, several increases from an initial number of $0$ (or close to $0$) to considerably higher numbers can be triggered for the web-based method \tool{GSEA}, which can be of considerable value to a researcher. \\
\textit{\textbf{True sample labels:}} 
The number of DEGS can be tweaked for all considered methods apart from \tool{DAVID} and \tool{clusterProfiler}'s ORA. For the latter, however, the number of DEGS, exceeding $300$, is already substantial. For \tool{GOSeq} and \tool{PADOG}, an increase of $1$ DEGS is induced through the choice of the method to transform the RNA-Seq data and the choice of the gene set database KEGG, respectively. Given the initial number of $2$ and $1$ DEGS, respectively, this increase is high in relative terms. Considerable `absolute' increases can be achieved for the three GSEA-based methods. For \tool{GSEAPreranked} and \tool{clusterProfiler}'s GSEA, respectively, a notable increase from $0$ is obtained solely through weighting each gene equally in the computation of the enrichment score. The highest increase among all methods is obtained for the web-based application \tool{GSEA} through the exploitation of uncertainty in several analytical choices in data preprocessing and the parameters. Thereby, the strongest increase is induced through the modification of the weighting pattern of the genes in the computation of the enrichment score.

\subsubsection{Results for goal 2: minimise adjusted $p$-value}

\begin{figure}
    \centering    \includegraphics[scale = 0.6]{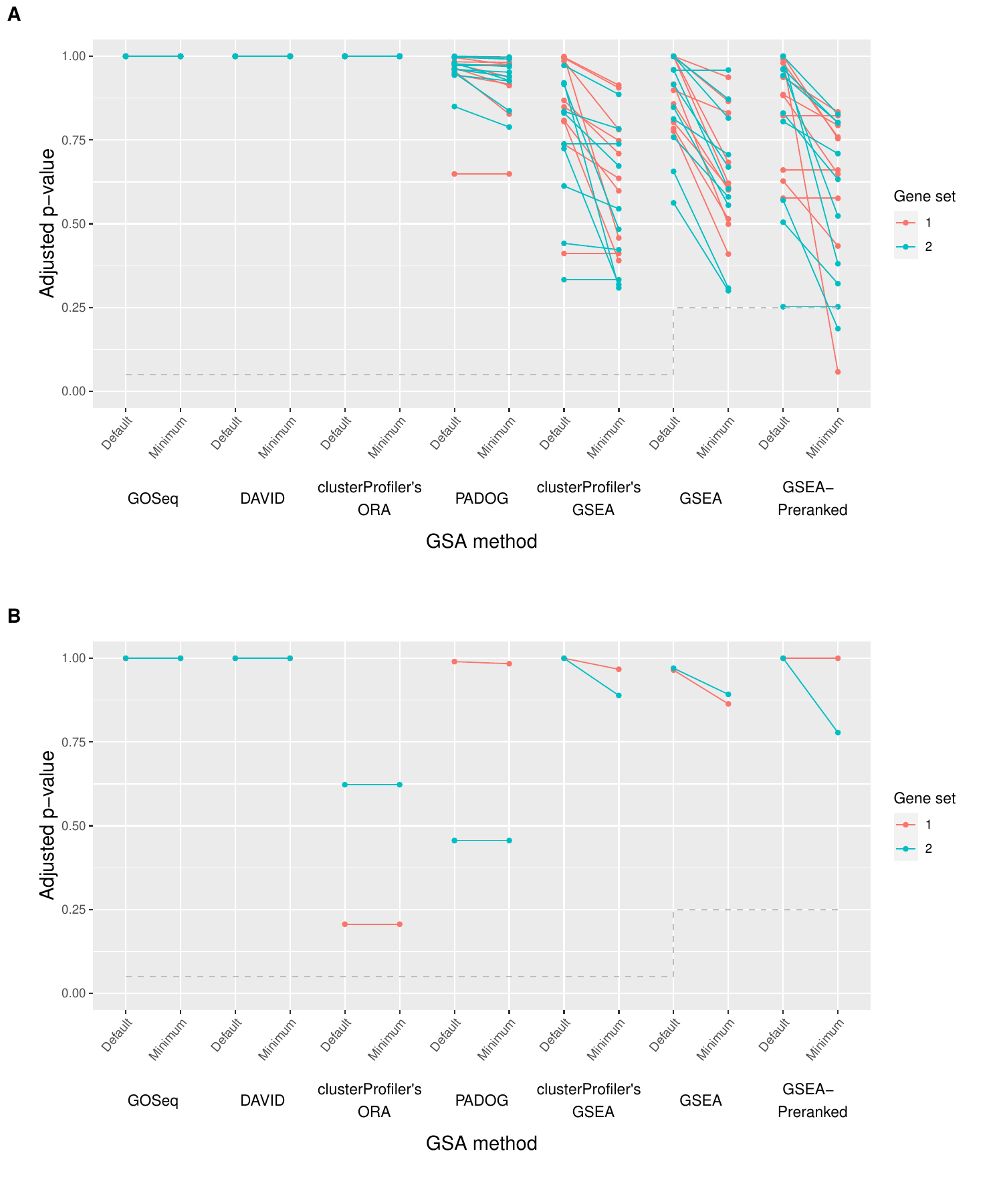}
    \caption{\small Goal 2: The optimised adjusted $p$-values (/$q$-values) in the Bottomly data set (`Minimum'), obtained through the exploitation of uncertainty, are compared to the corresponding values resulting from the default analytical choices (`Default'). Note that for the web-based applications \tool{GSEA} and \tool{GSEAPreranked}, the $q$-value is used to assess differential enrichment instead of the adjusted $p$-value. For each optimisation process, the associated optimised and the default adjusted $p$-value (/$q$-value) are connected through a line. (A) presents the results for the ten random permutations and (B) for true sample labels. On the $x$-axis, the individual methods investigated in the context of goal 2 are displayed. The results for gene set~1 are shown in red and those for gene set~2 in blue. The dashed grey line indicates the significance threshold for each method below which a gene set is considered differentially enriched.}
    \label{fig_results_padj_supp}
\end{figure}

\textit{\textbf{Random sample label permutations:}} 
For \tool{GOSeq}, \tool{DAVID}, and \tool{clusterProfiler}'s GSEA, we are not able to induce any over-optimistic results through the exploitation of data preprocessing and parameter uncertainty in the context of goal 2. In particular, for these three methods, the adjusted $p$-values of both gene sets amount to $1$ before and after the exploitation of uncertainty in all of the permutations. For \tool{PADOG}, the adjusted $p$-values resulting from the default choices are close to $1$ in the vast majority of permutations. While they can be tweaked for several permutations, particularly through the choice of the pre-filtering method and the method to transform the RNA-Seq data, the effect is mostly negligible, resulting in tweaked adjusted $p$-values that are still close to $1$. \\
For \tool{clusterProfiler}'s GSEA, web-based \tool{GSEA} and \tool{GSEAPreranked}, we observe moderate to notable decreases in the adjusted $p$-values (/$q$-values) for both gene sets and in the majority of permutations. While for the former two methods, neither of them transforms an initially non-significant adjusted $p$-value (/$q$-value) into a significant one, a significant $q$-value of enrichment is obtained in one permutation for gene set~1 and gene set~2, respectively, for \tool{GSEAPreranked}. For the latter gene set particularly, the choice of the DE method as part of data preprocessing and the modification of the weighting pattern of the genes in the computation of the enrichment score tweak the $q$-value from $0.99$ to $0.06$.\\ 
\textit{\textbf{True sample labels:}} 
For all of the considered methods and the respective two gene sets, we observe negligible to at most moderate decreases in the respective adjusted $p$-values (/$q$-values), such that the tweaked adjusted $p$-values (/$q$-values) are far from the corresponding significance thresholds.

\subsubsection{Results for goal 3: minimise rank of a specific gene set}

\begin{figure}
    \centering    \includegraphics[scale = 0.65]{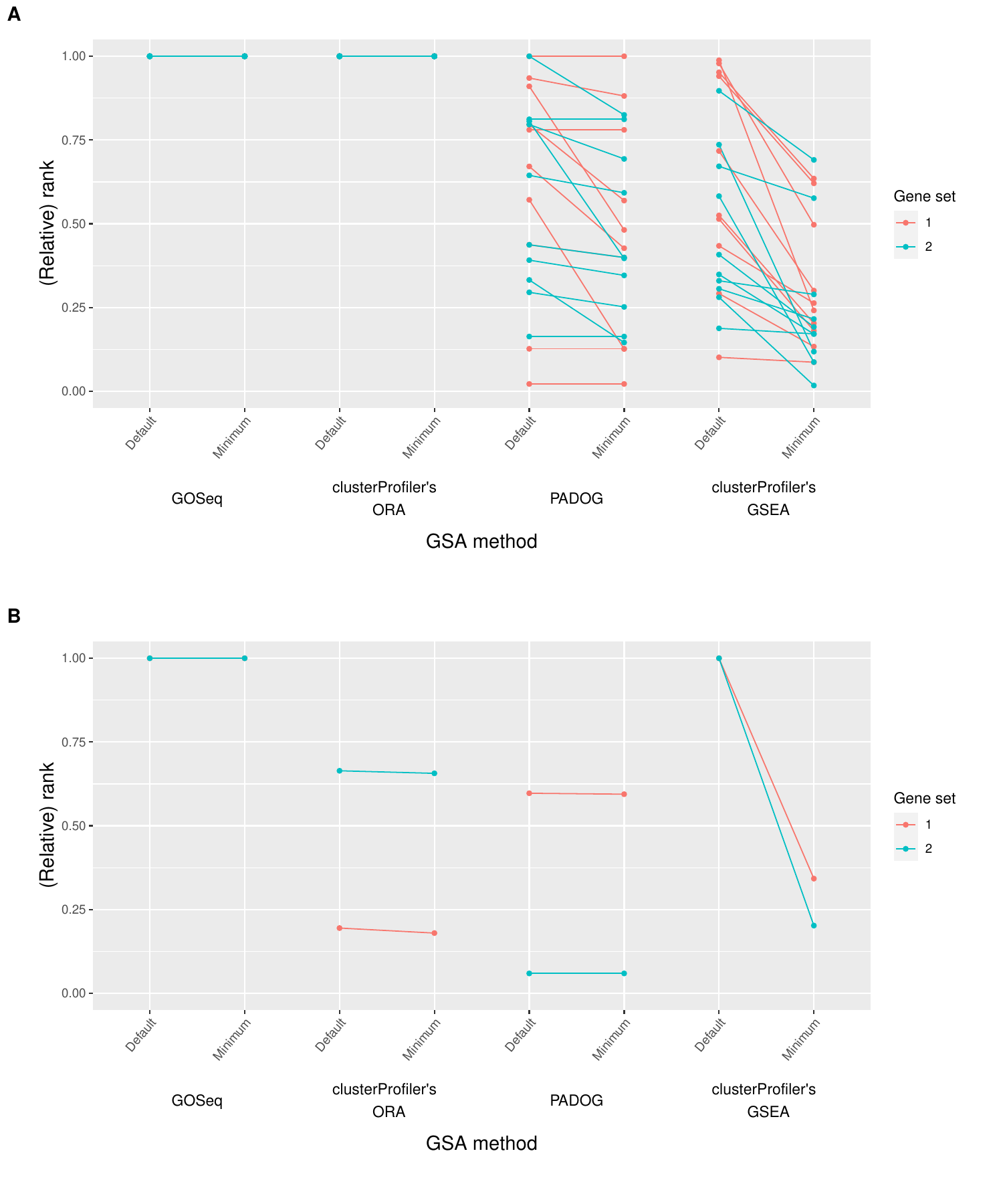}
    \caption{\small Goal 3: The (relative) ranks (`Minimum') in the Bottomly data set, obtained through the exploitation of uncertainty, are compared to the corresponding values resulting from the default analytical choices (`Default'). For each optimisation process, the associated optimised and the default rank are connected through a line. (A) presents the results for the ten random permutations and (B) for true sample labels. On the $x$-axis, the individual methods investigated in the context of goal 3 are displayed. The results for gene set~1 are shown in red and those for gene set~2 in blue.}
    \label{fig_results_relrank_supp}
\end{figure}

\textit{\textbf{Random sample label permutations:}} 
Similar to the study on the Pickrell data set in the main document, we observe no over-optimistic results for \tool{GOSeq} and \tool{clusterProfiler}'s ORA in the context of goal 3. In particular, the relative ranks in all permutations amount to $1$ before and after the exploitation of uncertainty. This is consistent with the observation on the adjusted $p$-values of $1$ observed for goal 2 (see Figure~\ref{fig_results_padj_supp}). For \tool{PADOG}, we, again, observe moderate decreases in the ranks for the majority of permutations for both gene sets, all of which are triggered through the choice of pre-filtering or the method to transform the RNA-Seq data. As with the Pickrell data set, the ranks of both gene sets are moderately low for many of the permutations, therefore indicating a moderately high relevance to the condition of interest compared to the remaining gene sets. When inspecting the adjusted $p$-values, however, we observe that the level of adjusted $p$-values of both gene sets are generally high across all permutations. This indicates that the majority of the remaining gene sets in the GSA results are even higher (i.e. close to $1$).\\ 
For \tool{clusterProfiler}'s GSEA, we observe that the relative ranks of both gene sets differ considerably between the different permutations before exploiting any uncertainties. In the majority of permutations, further decreases of moderate to notable magnitude can be obtained through various combinations of the analytical choices. However, for some of the permutations, we also observe that the relative rank and the corresponding adjusted $p$-value can differ considerably in magnitude. For instance, for gene set~2, we observe a decrease in the relative rank from $0.41$ to $0.19$ in one permutation. However, the corresponding adjusted $p$-values amount to $0.83$ and $0.92$, respectively. This indicates that the adjusted $p$-values of the remaining gene sets in the results table increase even more strongly through the exploitation of uncertainty than the adjusted $p$-value of gene set~2.  \\
\textit{\textbf{True sample labels:}} \\
We observe no to negligible decreases in the relative ranks of both gene sets for \tool{GOSeq}, \tool{clusterProfiler}'s ORA, and \tool{PADOG}. For \tool{clusterProfiler}'s GSEA, on the other hand, the ranks can be tweaked notably, namely from $1$ to $0.34$ and from $1$ to $0.20$ for gene sets 1 and 2, respectively. This implies a substantial increase in the `relative' relevance of both gene sets to the condition of interest compared to the remaining genes in the corresponding results tables. However, the corresponding adjusted $p$-values remain high even after the exploitation of uncertainty. For instance, the tweaked rank of $0.2$ of gene set~2 corresponds to an adjusted $p$-value of $0.98$, indicating that the adjusted $p$-values of the remaining gene sets in the result table are even closer to $1$. A similar observation is made for gene set~1.

\end{document}